# BioScore: A Foundational Scoring Function For Diverse Biomolecular Complexes

Yuchen Zhu[#1], Jihong Chen[#1], Yitong Li[#1], Xiaomin Fang[2], Xianbin Ye[2], Jingzhou He[2], Xujun Zhang[1], Jingxuan Ge[1], Chao Shen[3], Xiaonan Zhang[2*], Tingjun Hou[1,3,4*], Chang-Yu Hsieh[1,3,4*]

[1] College of Pharmaceutical Sciences, Zhejiang University, Hangzhou, 310058, China

[2] Baidu Online Network Technology (Beijing) Co., Ltd. Beijing, 100085, China

[3] The First Affiliated Hospital, College of Medicine, Zhejiang University, Hangzhou, 310058, China

[4] Zhejiang Provincial Key Laboratory for Intelligent Drug Discovery and Development, Jinhua, 321016, Zhejiang, China

    Chang-Yu Hsieh

    E-mail: kimhsieh@zju.edu.cn

    Tingjun Hou

    E-mail: tingjunhou@zju.edu.cn

    Xiaonan Zhang

    E-mail: zhangxiaonan@baidu.com

#These authors contributed equally.

# Abstract

Structural assessment of biomolecular complexes is essential for translating molecular modeling into functional insights in life science, impacting our understanding of biological mechanisms and informing drug discovery. However, existing structure-based scoring functions are typically system-specific, lacking generalizability across the diverse chemical and structural landscape of biomolecular complexes. We introduce BioScore, a foundational scoring function for assessing biomolecular structures and interactions. BioScore addresses three critical challenges—data sparsity, cross-system representation, and task compatibility—through a unified dual-scale geometric graph representation learning with specifically designed structure assessment and affinity prediction modules. This framework enables BioScore to support a wide array of tasks, including affinity prediction, conformation ranking, and structure-based virtual screening. Across 16 benchmark tasks spanning proteins, nucleic acids, small molecules, and carbohydrates etc, BioScore consistently outperforms or matches 70 traditional tools and deep learning methods. This rigorous benchmarking procedure also features our newly proposed PPI Benchmark—a comprehensive benchmark system for protein–protein complex scoring. BioScore's near universal applicability is demonstrated in three aspects: (1) pre-training on mixed-structure data yields up to 40% improvement in protein–protein affinity prediction and over 90% increase in Pearson correlation for antigen–antibody binding; (2) cross-system generalizability enables zero- and few-shot prediction for nucleic acid interacting with molecule and protein with improvements of 15% and 71% in Pearson correlation, respectively; and (3) its unified representation successfully characterizes chemically challenging systems such as cyclic peptides, achieving over 60% Pearson gain in affinity prediction. BioScore thus establishes a scalable, robust, and generalizable framework for structural assessment tasks across a broad spectrum of biomolecular systems.

# Introduction

In recent years, artificial intelligence (AI) has achieved transformative progress across a range of domains, from natural language processing and image recognition to biomedical research. AI models have evolved from specialized, task-specific tools into general-purpose systems with cross-domain capabilities[1–4]. Within the growing movement of "AI for Science," AI has been increasingly deployed to tackle grand scientific challenges, particularly in the life sciences and drug discovery[5,6]. Central to this effort is the pursuit of models with robust generalizability and strong zero-shot or few-shot learning capabilities across a wide spectrum of downstream tasks[7–9]. In the study of biomolecular complexes, this has catalyzed the development of large-scale foundation models aimed at bridging diverse molecular classes, with the goal of developing a universal representation and interpretation across complex biological systems[10,11].

Several landmark studies have laid the groundwork for distinct foundational frameworks for biomolecular modeling. The Evo model series[12,13] applies language modeling techniques to genomic sequences, effectively capturing diverse modalities—DNA, RNA, and proteins—through a unified lens. In structural modeling, AlphaFold3[14] and ESMFold[15] are excellent examples of how to build universal structure predictors extending across increasingly diverse biomolecular systems and bridge the modality gap between structure and sequence, respectively. Kong et al. introduced an equivariant Transformer architecture capable of modeling the hierarchical organization and physical interactions among molecular entities[14]. More recently, Fang et al. presented ATOMICA, a generalist model for characterizing interaction interfaces spanning small molecules, metal ions, amino acids, and nucleic acids[16]. Together, these efforts outline a grand scheme to unify structure and semantics that can universally characterize a variety of biomolecules and their interactions.

For structural assessment tasks in life science, the AlphaFold series stands as a milestone achievement[14]. Yet despite their remarkable capabilities in structure prediction, recent studies reveal a key limitation: AlphaFold3 exhibits poor performance in ranking binding affinities, making it inadequate for distinguishing active from inactive compounds[17]. This shortcoming underscores a broader challenge in the field[18–21]: while biomolecular structure prediction has made tremendous progress[14,22–24], the systematic evaluation of inter-molecular interactions currently lacks a transformative solution of comparable impact. As most essential biological processes critically depend on the strength of biomolecular interactions, assessing binding affinity and the stability of binding conformations is a critical step in elucidating the mechanisms of life and advancing drug design[25]. However, experimentally resolving the 3D structures of biomolecular complexes is often costly and time-consuming[26], it has not been easy to directly obtain the primary binding mode along with binding affinity of biomolecular complexes at large scale.

To address this gap, scientists leveraged molecular docking techqniue as early as the 1990s[27], aiming to computationally model complex structures with a score function in drug design. With technological advancements, the role of scoring functions has expanded beyond simply evaluating docking poses. Ideally, a scoring function should naturally provide affinity prediction and screening functionalities, enabling not only the assessment of inter-molecular binding strengths but also supporting tasks such as virtual screening and lead compound optimization. In the fields of life sciences and drug design, scoring functions are becoming a vital bridge between structure prediction and functional interpretation[28].

Historically, scoring functions fall into two broad categories: physics-based or empirical models that approximate binding affinities via pre-defined energy terms, and statistical models that extract potential functions from known structures[29,30]. Both approaches rely on pre-defined functional forms and assume simple relationships between input features and binding affinity, limiting their flexibility and generalizability. Deep learning offers a powerful alternative[31]. By learning complex, nonlinear mappings from structural data to affinities, AI-based scoring functions have shown promise[32]. However, most current models are narrowly trained on protein–small molecule datasets and exhibit limited transferability to systems involving more complex molecular scaffolds—macrocycles and natural products—as well as biomolecules such as nucleic acids, peptides, and antibodies etc. This narrow scope constrains their applicability across the full landscape of therapeutic targets, which increasingly includes protein–protein and protein–nucleic acid interactions—key players in immune signaling, transcriptional regulation, and disease pathogenesis[25,33,34].

This gap raises an important question: in the era of foundation models, can we develop a universal scoring function for biomolecular complexes? Several key challenges must be addressed. First, data sparsity: outside of protein–small molecule systems, there is a dearth of labeled structural data with experimentally measured affinities[35–37]. Second, representational complexity: distinct biomolecular systems exhibit diverse interaction modes and physicochemical properties, complicating the design of generalizable feature representations. Third, functional tradeoffs: while an ideal scoring function should support scoring, ranking, docking, and screening, current models are often optimized for a single task at the expense of others—a problem that becomes more acute when extending to diverse biomolecular systems.

Despite these challenges, a universal scoring function remains both physically plausible and practically valuable. Fundamentally, biomolecules—whether proteins, nucleic acids, or small molecules—share a common chemical basis, being composed of atoms from a limited set of elements. This opens the door to atom-level representations that transcend molecular species. Moreover, the physical forces that govern molecular interactions—electrostatics, van der Waals forces, hydrogen bonding etc—are universal. These shared principles support the theoretical foundation for a generalizable evaluation framework.

Importantly, a foundational scoring function would fill a critical gap in the modeling pipeline. While much attention has focused on the unified molecular encoders, scoring functions represent the final, decision-making step in structure-based tasks such as virtual screening and affinity optimization. For instance, cyclic peptides—an emerging class of therapeutics with unique macrocyclic features—are gaining traction in drug development. Yet current scoring functions, predominantly tailored for protein–small molecule systems, are ill-equipped to handle the modeling of highly flexible macrocycles and related structures. A universal scoring function could bridge this gap, enabling broader and more accurate structural assessments across diverse biological systems.

Here, we present BioScore—the first foundational scoring function designed to assess binding phenomanon across a broad range of biomolecular complexes (Fig. 1a). BioScore's technical highlights can be summarized as follows. (1) Representation: It departs from traditional atom/block discretizations by introducing interface-masking encodings and distance-aware edge construction, capturing dual-scale atomic and block-level features. This approach enables coarse-grained yet expressive representations applicable across molecular classes of various structural complexity. (2) Scoring Methodology: BioScore proposes a new structural assement score that incorporates a learned statistical potential (via a mixture of density network, MDN) and a newly defined interaction-edge-aware score. (3) Training strategy: A pretraining–fine-tuning workflow prioritizing large-scale structural learning at the pretraining stage balances performance across heterogeneous tasks, enabling plug-and-play adaptation to specific molecular systems. During the development of BioScore, we recognized a lack of suitable benchmarks to evaluate scoring functions for protein–protein interactions (PPIs). To this end, we constructed a newly proposed PPI Benchmark, the first comprehensive dataset for evaluating the full-spectrum capabilities of scoring functions in protein–protein complexes. BioScore is evaluated across 16 tasks—including classical systems (protein–small molecule, protein–protein, protein–nucleic acid, nucleic acid–small molecule) and specialized systems (e.g., cyclic peptides, macrocycles, carbohydrates, antibody–antigen, and peptide–MHC complexes). Against 70 traditional and domain-specific deep learning baselines (Table S1), BioScore consistently achieves state-of-the-art (SOTA) or near SOTA performance across domains (Table S2).

In this study, three key findings demonstrate the generalizability of BioScore. (1) Cross-domain pretraining significantly improves the overall performance, including 40% gains in protein–protein affinity prediction and over 90% relative improvement in Pearson correlation for antibody–antigen binding. (2) Structural transfer learning enables accurate zero-shot prediction on nucleic acid–small molecule systems and boosts few-shot performance on protein–nucleic acid complexes by 15% and 71% (Pearson correlation) over prior SOTA, respectively. (3) Universal representation learning enhances performance on challenging molecular classes that lie at the blurring boundary of the traditional notion of small molecules, such as macrocycles and some carbohydrates. For example, BioScore delivers over 60% relative gain in correlation for protein–cyclic peptide interactions compared to baseline methods. As the first universal scoring function tailored for diverse biomolecular complexes, BioScore offers a unified, extensible framework for cross-system evaluation, providing a robust foundation for mechanistic analysis and drug discovery across complex biological systems. As the landscape of potential therapeutic targets and drug modalities expands beyond traditional proteins and Rule-of-Five small molecules, BioScore fills a critical gap by enabling rigorous evaluation of prospective drug candidates.

# Results

## Design Principles of BioScore

Most existing deep learning–based scoring functions rely on domain-specific encoding strategies and task-specialized designs, limiting their ability to generalize across different types of biomolecular complexes[38,39]. In contrast, BioScore integrates the principles of statistical potentials with a unified geometric representation framework, aiming to strike a balance between generality and effectiveness.

First, to enhance the cross-system consistency of complex representations, BioScore integrates dual-scale information from both atoms and building blocks such as residues and nucleobases (Figure 1b). It constructs an all-atom geometric graph at the atomic level while incorporating domain-specific block-level information. This unified geometric graph structure allows the model to accommodate various biomolecular systems, including proteins, nucleic acids, and molecules of varying degrees of structural complexity. Additionally, a feature extraction module based on a general equivariant Transformer (Figure 1f) is employed to capture both intra-domain and cross-domain information across different biomolecular complexes.

While explicitly adding inter-molecular interaction edges within complexes is a common practice in representation learning—especially emphasized in prior work on general biomolecular representation[16,40,41]—it induces overfitting when applied in scoring functions based on statistical potentials (Figure S1), ultimately compromising the modeling of distance distributions. To address this challenge, we propose a interface-masking encoding strategy that deliberately masks inter-molecular edges and retains only intra-molecular edges, thereby enhancing the model's capacity to perceive and learn true spatial distances. This design is crucial for constructing a foundational scoring function applicable to general biomolecular complexes. Ablation experiments demonstrate that the introduction of the interface-masking strategy significantly improves docking and screening performance across various structural complexity and chemical diversity, including protein–ligand (small molecule), protein–protein complexes (Tables S3–S4).

In complex system modeling, BioScore further introduces a distance-threshold-based edge construction strategy. This approach builds intra-molecular and inter-molecular edges by applying tailored distance cutoffs for different biomolecules and edge types: 10 Å for proteins/nucleic acids[38], 2 Å for small molecules[42], and 8 Å for complex interaction edges. These thresholds are determined by empirical experiences as well as extensive testing. We settle for this rule-based method for now, instead of a dynamical, self-adapted threshold setting to streamline the numerical processing. Compared to the common K-nearest neighbor (KNN) methods or enumeration strategies, this distance-cutoff strategy better characterizes the critical edge count—which is essential for the downstream dual-tower output module—while significantly improving computational efficiency. Specifically, compared to our previous model RTMScore[38] developed for protein-ligand (PLI) systems, BioScore achieves a 15-fold improvement in computational speed, as shown in Table S5 (We have also included statistics for several important scoring functions used in our study, which will be discussed in detail in later sections).

Finally, BioScore employs a dual-tower scoring architecture designed to address two task categories: docking/screening and scoring/ranking (Figure 1e). In the docking and screening pathway, the model constructs statistical-potential-based scoring terms based on the principles of inverse Boltzmann distribution, and introduces an interaction-edge-count-based confidence term as an auxiliary signal to enhance the model's ability to distinguish true binding conformations. This strategy significantly improves docking and screening performance across multiple complex systems (Tables S3–S4). In the scoring and ranking pathway, the model fits nonlinear relationships between interacting atom pairs and binding free energies using neural networks, outputs the mean score as the primary prediction, and incorporates an edge-count-aware confidence term to improve the characterization of binding affinity strength. Together with our pretraining–fine-tuning strategy (Figures 1d and 1e), BioScore is capable of learning structural features from diverse biomolecular systems collected from multiple datasets (Table S6) and adapting to task objectives across different molecular species. During pretraining, BioScore specifically trains the MDN module by leveraging pairwise distance distributions derived from experimental complex configurations. This enables BioScore to capture general structural features by minimizing the negative log-likelihood of the distance distributions. In the subsequent fine-tuning phase, the model undergoes task-specific fine-tuning on each type of complex, such as protein–protein interactions, molecular binding affinity, or other relevant supervised tasks, while combining multiple loss components to adapt its learned representations to specific downstream objectives.For a detailed description of BioScore, please refer to the Methods section.

## A New Comprehensive Benchmark for Protein–Protein Interactions

One of the key challenges in developing a universal scoring function is to establish systematic evaluation across different molecular systems. At present, only the protein–small molecule system has some widely accepted scoring benchmarks—such as CASF-2016[43]—while other biomolecular systems, particularly protein–protein complexes, lack a standardized benchmark dataset. This gap leads to significant inconsistencies in evaluation criteria, data sources, and capability coverage across different methods, hindering fair and consistent comparisons.

To address the aforementioned challenges, we constructed a benchmark dataset for structural assesments of protein–protein complexes (PPI Benchmark) based on the PDBbind v2020 database[44], covering three core tasks: scoring, docking, and screening. The benchmark design is inspired by CASF-2016, while ensuring data quality and source consistency.

We first selected 177 representative protein–protein complexes, then further filtered 79 complexes considering the diversity of chain lengths to construct the foundational dataset (Figures 2a, 2b). Based on this, we further three task-specific subsets: (1) Scoring benchmark: comprising the 79 native complexes with experimental binding affinity annotations, used to evaluate binding affinity prediction performance (Figure 2b); (2) Docking benchmark: for each complex, decoy conformations were generated using ZDOCK[45], and 7,979 conformations were selected based on DockQ scores to ensure balanced quality distributions, used to assess the model's ability to distinguish native-like from non-native conformations (Figure 2c); (3) Screening benchmark: for each of the 79 receptor–ligand pairs, cross-docking was performed, and 613,900 conformations were obtained through K-means clustering and sampling, used to evaluate the model's ability to discriminate active from inactive ligands (Figure 2d).

Compared to existing benchmarks, our newly proposed PPI Benchmark represents the first multidimensional evaluation framework for protein–protein complexes built from a unified data source, providing an objective and reproducible foundation for comprehensive scoring function evaluation. Detailed construction procedures and evaluation metrics are described in the Supplementary Information.

## Comprehensive Evaluation on Protein–Protein Interactions

We first systematically evaluated the three core capabilities of BioScore on the custom-built PPI Benchmark and compared its performance against various deep learning models (GNN-DOVE[46], DProQA[47], GET[41], MINT[48]) and classical methods (ZRANK2[49], VoroMQA[50]) (Table 1). BioScore ranks first in 9 out of 13 metrics, highlighting its superior overall performance. Detailed descriptions of the baseline models are provided in the Supplementary Information.

In the scoring task, BioScore ranked second only to MINT, a model recently developed by Ullanat et al., which was pretrained on 96 million sequence pairs from STRING-DB[51], whereas our model was trained on only ~1,700 structural data points (a detailed description of the training and testing data across all evaluation tasks is provided in the Methods section). In the docking task, BioScore significantly outperformed other methods in terms of the success rate for identifying native and high-quality conformations (Table 1, Figures 3a and 3b) and demonstrated strong robustness under diverse decoy distributions. In the screening task, BioScore was the only method that exhibited effective discrimination capability on our constructed screening benchmark (Figure 3c), further validating the advantages of statistical-potential-based scoring approaches and their applicability to protein–protein complexes when appropriately adapted. Moreover, while existing models tend to excel in specific functionalities, they often lack generality. For instance, although GET and MINT perform well in scoring tasks, they are limited in docking and screening tasks, while VoroMQA shows docking capabilities but lacks affinity predictive power. BioScore is the only model that simultaneously integrates all three functionalities among tested methods on the newly proposed PPI benchmark.

To further evaluate BioScore's generalizability, we applied it to two more challenging tasks: antigen–antibody affinity prediction and peptide–MHC-I screening. Antibodies recognize antigens through their complementarity-determining regions (CDRs) in the variable domains, which are highly variable in both structure and sequence and exhibit low evolutionary conservation, making binding affinity prediction particularly difficult[52]. We collected 272 complex structures with affinity labels from the SAbDab database[53] and divided them into training, validation, and test sets (in a 3:1:6 ratio) while controlling for data leakage. The results show that BioScore significantly outperforms GET and MINT in this task (Figure 3d, Table S7).

In the peptide–MHC-I task, we previously constructed a screening benchmark comprising 11 alleles based on the pipeline proposed by ITN[54]. Under the zero-shot condition—without fine-tuning on any task-specific data—BioScore outperformed all baseline models in both AUROC and AUPR metrics (Figure 3e), demonstrating strong cross-domain generalization capabilities.

## Strong Generalization and Robustness in Protein–Ligand Scoring Tasks

Prediction of protein–small molecule interactions (PLI) is of great importance in the early stages of drug discovery, serving as a key component in molecular docking and virtual screening workflows[32]. Given the relative abundance of high-quality data in this domain (for example, nearly 20,000 PLI entries are available in PDBbind v2020[44]), PLI prediction has become a major focus for both traditional scoring function development and deep learning model research, leading to the establishment of widely used benchmark datasets such as CASF-2016[43].

Our previously developed model, RTMScore[38], demonstrated strong performance in PLI tasks but still faced three major limitations: (1) it required enumerating all possible interaction edges between proteins and small molecules, leading to low computational efficiency; (2) it adopted separate encoding strategies for proteins (receptor) and small molecules (ligand), preventing effective information exchange across the two encoders and resulintg in limited generalizability; and (3) it primarily supported docking and screening tasks, lacking in scoring and ranking capabilities. To address these challenges, we subsequently introduced GenScore[55], which incorporated a correlation loss based on binding affinity into the loss function, thereby improving ranking capabilities. However, GenScore still could not reliably predict absolute binding affinity values. These limitations have been systematically addressed in BioScore.

On the CASF-2016 benchmark, we constructed the training data exclusively from PDBbind, without introducing any additional data augmentation. The results demonstrate that BioScore outperforms multiple mainstream methods—including many first-tier, SOTA methods such as RTMScore and GenScore—across the scoring, ranking, and screening tasks (Figure 4, Table S8). In the docking task, BioScore ranks third, achieving comparable performance to GenScore and RTMScore, indicating that it maintains strong structural discrimination capabilities while preserving its modeling generality.

To further validate its generalization performance, we evaluated BioScore on two external screening benchmarks, DUD-E[56] and DEKOIS2.0[57], and compared its performance with EquiScore[58], a SOTA screening model. The results indicate that on DUD-E, BioScore ranks first across all three enrichment factor metrics ($EF_{1\%}$, $EF_{5\%}$, $EF_{10\%}$), and second in the BEDROC metric, slightly behind RTMScore (Figure S2). On DEKOIS2.0, while BioScore trails behind EquiScore and RTMScore, it consistently ranks within the top three across most metrics (Figure S3), demonstrating robust screening capability and generalization across datasets.

Finally, we selected one protein–small molecule complex from our benchmark set, 3KR8, which corresponds to the catalytic domain of human tankyrase 2. Tankyrases are involved in fundamental cellular processes such as telomere homeostasis and Wnt signaling[59]. Figure 4e–f presents the residue–atom distance map and the corresponding energy contribution map calculated by BioScore for the 3KR8 complex, clearly demonstrating a strong correlation between spatial proximity and binding energy contribution. Furthermore, based on previous structural and biochemical studies of this complex[60], we highlighted several key amino acid residues that are known to drive the interaction with the ligand (Figure 4g). These residues show excellent agreement with those identified as high contributors in the BioScore-derived energy map, indicating that BioScore effectively captures biophysically meaningful interaction patterns relevant for inhibitor binding.

## Cross-Domain Pre-training Enhances BioScore's Perfermence

In the aforementioned PLI and PPI benchmark evaluations, BioScore outperformed domain-specific methods despite not relying on any task-specific data augmentation strategies. Unlike previous approaches that improve performance through manually constructed positive/negative samples or by training model with an increased task-specific data volumes, BioScore's enhancement stems from its universal biomolecular complex modeling capability—achieved through cross-domain mixed training on heterogeneous complex systems. This type of training mechanism is only feasible within a model architecture that supports unified representation scheme and task adaptation capability.

As previously discussed, deep learning–based scoring functions commonly face the challenge of data sparsity. We posit that an ideal foundational scoring function should possess the ability to extract shared structural information across diverse biomolecular complexes, thereby improving performance on any individual system. Following this principle, BioScore is first pretrained on a combined dataset of PLI and PPI structural data, and then fine-tuned on labeled data for each specific system. To systematically validate the effectiveness of this strategy, we designed control experiments comparing performance under two conditions: pretraining on single-domain data versus pretraining on cross-domain PLI and PPI mixed data, and evaluated their impact on both PPI and PLI tasks (Tables S9–S12).

In the PPI task (Table S9), cross-domain training led to substantial improvements: across the three affinity prediction metrics, the average performance increase was 43.05%, with the Pearson correlation coefficient improving by as much as 109.09%. For the four screening capability metrics, the average improvement reached 42.17%, and the Top-1% success rate increased by 71.47%. These results indicate that despite differences in molecular types and binding modes between PLI and PPI systems, the physical and geometric patterns embedded in their 3D structures exhibit strong transferability when formulated in an appropriate framework. The inclusion of PLI data effectively compensates for the limited data availability in PPI systems and other cases (to be elucidated in subsequent sections). A similar trend was observed in the more challenging task of antigen–antibody binding affinity prediction. The incorporation of PLI pretraining data led to an average improvement of 37.27% across three key affinity prediction metrics, with the Pearson correlation showing a substantial gain of 91.67% (Table S10).

In the PLI task (Table S11), incorporating PPI data also had a positive impact on ranking capabilities, with an average improvement of 16.91%. Performance on scoring, docking and screening metrics remained stable, indicating that BioScore is robust in learning from heterogeneous structural information, particularly in data-rich scenarios.

Notably, this cross-domain training strategy also proves effective for systems in specialized chemical spaces (Table S12), such as protein–cyclic peptide complexes, demonstrating broad transfer potential. We will present further investigations on these specific cases below.

We also explored the combination of PLI and PPI data during the fine-tuning stage. However, the results (Table S13) indicate that while this approach still outperforms single-domain fine-tuning, it is slightly less effective than our current strategy—performing cross-domain mixing during pretraining and using only target-domain data during fine-tuning. We hypothesize that this is because structural modeling exhibits stronger generality, whereas label fitting is more sensitive to data distribution. Based on this strategy, BioScore maintains its generalization capability while also offering "plug-and-play" flexibility and efficiency for specific task scenarios, aligning with the dual demands for resource efficiency (e.g. training data volume) and accuracy in practical applications.

# Scoring Nucleic Acids: From Protein–Nucleic Acid to Nucleic Acid–Ligand

In the previous section, we clearly demonstrated the benefits of cross-domain training for PPI and PLI tasks. A follow-up question naturally arises: can BioScore maintain its competitive performance for other biomolecular systems? Biomolecular complexes (relevant for life science and drug discovery) usually come from the four major categories: protein–ligand interactions (PLI), protein–protein interactions (PPI), protein–nucleic acid interactions (PNI), and nucleic acid-ligand interactions (NLI). Among them, protein–nucleic acid interactions (PNI) play a fundamental role in essential biological processes such as gene expression regulation, cellular function maintenance, and pathogen immunity, making them a critical focus in the drug target research[34]. However, compared to other biomolecular complex systems, affinity prediction for PNI poses greater challenges, primarily due to the high structural diversity and high structural flexibility inherent to nucleic acid molecules[61].

To systematically evaluate BioScore's scoring capability for PNI, we exclusively used 1,052 valid PNI entries from PDBbind. To simulate the data scarcity commonly encountered in the real-world PNI scenarios, we strictly limited the training data volume, using only 276 samples (split 3:1:6 for training, validation, and testing, as shown in Figures 5a and 5b). During pretraining, PLI and PPI structural data were combined with PNI data for joint training, followed by fine-tuning on PNI label data. The results (Figure 5c, Table S14) demonstrate that BioScore outperforms multiple representative graph-based baseline methods across all scoring metrics. Specifically, compared to the best-performing baseline, BioScore improves the Pearson correlation by 70.88%. Notably, in RMSE and MAE, BioScore achieves highly accurate predictions of binding affinity (Figure 5f), showcasing its strong capacity for affinity prediction even under extremely low-resource learning conditions.

We further extended our evaluation to the more challenging scenario of nucleic acid–ligand complexes. Among all biomolecular complex systems, data for NLI are the scarcest[41], with only 134 valid entries in PDBbind—less than 1% of the PLI dataset. Here, we treated these 134 NLI samples as a zero-shot test set and directly evaluated binding affinity prediction using the model fine-tuned solely on PNI data. Remarkably, without having seen any NLI data, BioScore achieved excellent zero-shot transfer performance, significantly outperforming all baseline models (Figures 5d and 5h, Table S15). These results demonstrate that BioScore effectively captures shared structural and interaction patterns across systems and successfully generalizes to new scenarios with greater structural differences and extreme data scarcity.

It is worth noting that RNA–ligand interactions play essential roles in biological processes such as gene regulation and protein synthesis, serving as a key mechanistic basis for emerging research areas like antiviral drug development and non-coding RNA targeting[62]. To further validate the adaptability of BioScore, we selected this subtype as a representative case and conducted a detailed analysis of its affinity predictive power. We compared BioScore with traditional docking tools such as rDock, and also included MM/GBSA — a higher-accuracy molecular mechanics–based affinity estimation methods[63]. The experimental results (Figures 5e and 5h) demonstrate that even without explicit training on any NLI data, BioScore achieves prediction accuracy second only to the physics-grounded and computationally intensive MM/GBSA method and significantly outperforms all other methods, including rDock, which is specifically designed for RNA–ligand interactions.

# Generalization Over Challenging Regions in Chemical Space

The results from the preceding evaluations demonstrate the ability of a universal foundational scoring function to leverage information from different domains for performance enhancement in a specific domain, as well as its capacity for cross-domain generalization to complex, low-data, or even zero-shot scenarios. We further sought to investigate whether BioScore could be applied to certain molecular classes that have been challenging for prior methods for various technical reasons. Among these, cyclic peptides stand out as one of the most representative cases. Since the discovery of insulin in 1920[64], cyclic peptide drugs have demonstrated unique value across fields such as oncology, metabolic disorders, and infectious diseases, owing to their combination of small-molecule-like membrane permeability and antibody-level target selectivity[65]. Their molecular weights typically range from 500 to 5,000 Da, filling the gap between small molecules (<500 Da) and biologic drugs (>5,000 Da). From a chemical space perspective, we posit that cyclic peptides occupy an interesting region at the blurring boundary between small molecules and proteins (Figures 6a and 6b), implying that neither a purely small-molecule nor a purely protein-centric perspective is sufficient to fully characterize cyclic peptides with most existing data-driven methods. BioScore, as a scoring function for general biomolecular complexes, offers a unique opportunity to bridge small-molecule and protein perspectives for the study of cyclic peptides.

We previously constructed a protein–cyclic peptide benchmark dataset, CPSet, and evaluated the binding affinity prediction performance of various docking tools for protein–cyclic peptide complexes[66]. It is important to note that in both the current and previous evaluations, we excluded all samples containing macrocyclic structures (including cyclic peptides) from the PLI dataset, making this task a true zero-shot test. The results demonstrate that BioScore significantly outperforms all traditional scoring methods in terms of Pearson correlation coefficient (Figure 6c). Control experiments (Table S12) further show that BioScore only achieves its current optimal performance under the PLI and PPI mixed pretraining setup, highlighting its unique capacity to break through the limitations of single-domain perspectives and generalize to the specialized protein–cyclic peptide complex system.

Beyond cyclic peptides, non-peptidic macrocyclic drugs have also garnered increasing attention in recent years for drug design[67,68]. These molecules often exhibit natural product-like structures, typically do not conform to Lipinski's Rule of Five, yet hold unique potential for expanding chemical space in drug discovery (Figure 6d). We extracted non-peptidic macrocyclic data from PDBbind as a zero-shot test set to evaluate BioScore's scoring capability for protein–non-peptidic macrocycle complexes. Here, we again compared BioScore with baseline models such as GET, EGNN, and the traditional docking method AutoDock Vina. Without any additional fine-tuning, BioScore demonstrated performance far exceeding that of all baseline methods (Figures 6e and 6f, Table S16).

We further evaluated the applicability of BioScore to protein–carbohydrate systems. Such interactions play critical roles in fundamental biological processes, including cell recognition, signal transduction, and immune regulation[69]. However, their binding sites exhibit high structural diversity[70], making prediction much more challenging than in typical protein–small molecule systems (Figure 6g). Existing scoring functions for protein–carbohydrate interactions include CSM-carbohydrate[71], PCAPRED[72], and CSM-lig[73]. Notably, Nguyen et al., in the development of CSM-carbohydrate, manually curated a dataset of 370 protein–carbohydrate complexes with both structural and label information from resources such as ProCarbDB[74], ProCaff[75], and PCAPRED[72], and tested their model on an independent benchmark set of 43 protein–carbohydrate complexes. For a fair comparison, we fine-tuned BioScore exclusively on the same 370 labeled samples provided by CSM-carbohydrate, without using PLI labels from PDBbind to avoid potential data leakage issues. We then evaluated BioScore's performance on the same independent test set, and the results (Figure 6h, Table S17) demonstrate that BioScore, trained on the same label data, outperforms all existing methods and achieves SOTA performance.

In summary, BioScore demonstrates excellent performance across specialized complex systems, including cyclic peptides, non-peptidic macrocycles, and protein–carbohydrate interactions. These results highlight BioScore's capacity for structural transferability and task adaptability after overcoming the limitations of single-domain perspectives, underscoring its broad practical potential as a foundational scoring function.

## Discussion

In this study, we present BioScore—the first foundational scoring function specifically designed for structure-based assements of general biomolecular complexes. Unlike previous scoring methods developed for specific systems such as protein–ligand or protein–protein interactions, BioScore is applicable to a wide range of biomolecular complexes, offering broad compatibility for diverse structural inputs. While prior work on general molecular representation learning has enabled cross-system encoding, it has primarily focused on affinity scoring tasks, lacking structural evaluation capabilities such as docking and screening. BioScore represents a true technological advancements in this regard: by integrating interface-masking encoding, a distance-threshold-based edge construction strategy, and a dual-tower scoring design inspired by the physics-grounded method termed inverse Boltzmann distribution, it is the first model to unify scoring, ranking, docking, and screening functionalities within a single framework for general biomolecular complexes, addressing the longstanding challenge of balancing generality, task comprehensiveness and computational efficiency.

Across multiple regions of chemical space—including protein–protein and protein–ligand interactions—BioScore demonstrates leading performance on various standard evaluation metrics. Along the way, we also propose a new PPI benchmark suite that should facilitate future efforts in further developing scoring functions. Based on the comprehensive testing and analysis, BioScore exhibits exceptional generalization capability in challenging tasks under extreme data scarcity or zero-shot scenarios. For challenging cases such as antigen–antibody, peptide–MHC and nucleic acid–ligand complexes, BioScore outperforms many existing methods in either virtual screening or binding affinity prediction. This is made possible by BioScore's cross-domain knowledge transfer, enabled by its mixed-domain pretraining strategy. Furthermore, BioScore shows high adaptability across challenging molecular classes, such as cyclic peptide, non-peptidic macrocycle, and carbohydrate, filling a critical gap left by traditional models in handling fuzzy boundaries in chemical space.

Nevertheless, BioScore has room for improvement. Currently, the model does not fully leverage the vast amount of biomolecular complex data that contain only sequence and affinity labels, which are far more abundant than data with both structural information and labels. Future work could explore integrating multimodal techniques to embed sequence semantics into the structural modeling framework, thereby enhancing model performance in scenarios lacking structural data. Additionally, BioScore's performance in PLI screening tasks can definitely be improved, highlighting the need for more targeted data augmentation strategies to improve model discriminability and robustness. From an algorithmic perspective, traditional physics-based modeling methods that explicitly introduce functional form to model binding free energy remain valuable, and incorporating such physical priors into the current framework could further improve the interpretability and accuracy of the scoring function. Despite these exciting developments one can pursue in the near future, the current version of BioScore already offer strong performance and can be readily deployed in the real-world research and drug discovery projects. Furthermore, by integrating (and even fine tuning) BioScore with structure prediction and related binder design methods, such as AlphaFold3 and BindCraft etc, we see an exciting venue where structure prediction, assements and design all done in a self-consistent loop to greatly improve the task of binder designs.

# Method

## Preparation of the Training Dataset

BioScore adopts a pretraining–fine-tuning training strategy. The raw structural data and experimental binding affinity annotations of various biomolecular complexes used during training were primarily sourced from the widely adopted PDBbind database[44] (version 2020), a standard resource in the development of scoring functions. The data were preprocessed according to the following steps:

(1) Filtering

a. Given the significant differences in interaction patterns between covalently bounded protein-molecule complexes and general non-covalent systems, we preemptively removed all covalent complexes from the PLI dataset to prevent mixed inputs from interfering with model learning. The specific procedure was as follows: We first integrated three publicly available protein–covalent ligand structure databases—CovPDB[76], CovBinderInPDB[77], and CovalentInDB 2.0[78]—using the combination of the protein's UniProt ID and the normalized SMILES string of the small molecule (considering stereoisomers) as unique identifiers for each protein–covalent small molecule complex pair. For identical entries, only the first occurrence was retained, resulting in a total of 4,768 unique complex structures. Based on this, we applied the same "UniProt ID – normalized SMILES" matching criterion to identify and remove covalent complexes from the PDBbind (version 2020) PLI dataset, ultimately retaining 18,608 non-covalent PLI complexes for the subsequent construction of the training dataset.

b. Furthermore, considering the imbalance in structural and affinity annotation data distributions across different biomolecular complex systems, we established distinct filtering criteria tailored to the objectives of the two training stages to maximize the utilization of available data. For the pretraining stage, as no labels are required, we prioritized data diversity while applying only minimal structural constraints. Complexes were retained if they met the following conditions: (1) clear specification of interacting components; (2) absence of rare elements or non-canonical amino acids in the structure; and (3) a minimum of one heavy atom per component. For the fine-tuning stage, stricter criteria were applied based on conformation quality and affinity labels to ensure effective training: (1) structures must be determined by non-NMR methods with a resolution better than 2.5 Å; and (2) complexes must have explicit affinity annotations of type "Kd," "Ki," or "IC50," reported in units of "M" or convertible to "M." For PPIs, PNIs, and NLIs in PDBbind (with NLIs appearing only in test sets), given the small difference in retained data between the two filtering strategies, we uniformly adopted the stricter criteria for data quality considerations.

(2) Interaction Region Extraction

To ensure computational efficiency, we extracted interaction regions from the filtered data to replace full complex structures as model inputs. Based on differences in interaction patterns across biomolecular complex systems, we adopted system-specific extraction strategies. For PLIs, the ligand was retained in its entirety, while the receptor region was extracted using ProDy 2.4.1 by identifying all residues within 10 Å of the ligand structure. For PPIs, PNIs, and NLIs (the latter appearing only in test sets), both receptor and ligand regions were defined by distance thresholds, retaining only residues or atoms from both components that lie within 6 Å of each other.

(3) Deduplication

To prevent data leakage and ensure fair evaluation of model performance, all training data were deduplicated against benchmark test sets (CASF-2016, PPI Benchmark) by removing complexes with identical PDB IDs from the training and validation sets. Specifically, as recent studies (notably by Joseph Szymborski et al.[79]) have highlighted sequence similarity as an additional source of data leakage in PPI prediction tasks, we applied an additional sequence similarity–based deduplication step for all PPIs. Using the benchmark test sets as the query set and the pretraining/fine-tuning datasets as the reference set, we conducted sequence similarity searches via MMseqs2 15.6f452. Each complex's sequence was defined by concatenating the receptor and ligand chain sequences, with a sequence coverage threshold of 0.3. Based on the search results, we removed all reference set complexes with a minimum sequence similarity ⩾30% to any benchmark test complex. This stringent deduplication ensured the reliability of evaluation results. Furthermore, considering that the generality evaluation also involves specialized applications such as protein–cyclic peptide and protein–macrocycle systems, for all PLIs, in addition to PDB ID–based deduplication against CASF-2016, we also removed any complexes where the ligand was a macrocyclic molecule (identified by SMILES, defined as containing at least one ring with ⩾12 atoms).

(4) Dataset Splitting

Data that could not be effectively processed in the aforementioned steps were discarded. The retained data were randomly split into training and validation sets based on specific ratios. For PDBbind, following our previous work, we fixed the validation set size at 1,500 samples for PLIs. For PNIs, we reserved sufficient data for testing, adopting a 3:1:6 split for training, validation, and test sets. For all other cases without special instructions, a 9:1 split between training and validation sets was applied. Specifically, to address the impact of sequence similarity on PPI prediction tasks, PPIs were first clustered using MMseqs2 15.6f452 before random splitting. Two complexes were grouped into the same cluster if their sequence similarity exceeded 30% (with a sequence coverage threshold of 0.3). Clusters were then used as the splitting unit: 10% of the clusters were randomly selected as the validation set, and the remaining clusters formed the training set. All datasets used in the pretraining and fine-tuning stages are listed in Table S6.

## Model Performance Evaluation Description

To comprehensively evaluate the model's capabilities across different biomolecular complexes, we collected and constructed a series of test sets for extensive performance evaluation of BioScore.

(1) PPI Benchmark

Details of the PPI Benchmark construction are provided in the Supplementary Information due to space limitations.

(2) PLI Evaluation – Benchmark Dataset

CASF-2016 is a classic benchmark for protein–ligand interaction (PLI) scoring functions, constructed based on PDBbind PL. The core set comprises 285 active PLI complexes spanning 57 targets and is designed to evaluate four fundamental capabilities of models: scoring, ranking, docking, and screening.

(3) PLI Evaluation – External Test Sets

We adopted the widely used virtual screening benchmark datasets DEKOIS 2.0[57] and DUD-E[56] as external test sets for evaluating the screening capabilities of PLI models. DEKOIS 2.0 covers 81 different targets, with each target including 40 active ligands and 1,200 decoys. DUD-E includes 22,886 active ligands spanning 102 targets, with 50 decoys provided for each active ligand. For model evaluation, we generated at least 10 docking conformations per complex using Glide SP. All docking conformation data were sourced from our previous work (GenScore[55]).

(4) PPI Evaluation – Benchmark Dataset

Using the PPI Benchmark we constructed based on PDBbind PP as the benchmark, the core set comprises 79 active PPI complexes across different targets. This dataset is designed to evaluate model capabilities in scoring, docking, and screening. The specific construction process is described in detail above.

(5) PPI Evaluation – External Test Sets

We used an antigen–antibody test set constructed from the SAbDab database[53] as an external benchmark for PPI scoring evaluation. Complexes were filtered based on the following criteria: antigen type must be peptide/protein; antibody chains (heavy, light) and antigen chains must be clearly annotated; and binding affinity scores must be explicitly available. A total of 272 unique antigen–antibody interaction pairs were collected. We adopted a few-shot evaluation protocol: a small portion of the data was used for model fine-tuning, while the remaining data were reserved for evaluation. Specifically, the antigen–antibody complexes were clustered using MMseqs2 15.6f452, with a sequence similarity threshold of 80% (sequence coverage $\geq 0.3$). Clusters were used as the splitting unit and randomly divided into training/validation/test sets at a 3:1:6 ratio, resulting in 112/10/150 samples for training, validation, and testing, respectively. Additionally, we used the peptide–MHC virtual screening benchmark from our previous work (ITN[54]) as an external test set for PPI screening evaluation. This benchmark contains 42,459 unique complexes across 11 HLA class I targets, labeled for binary classification, with an overall positive-to-negative sample ratio of approximately 1:3. Considering that all peptide ligands in this dataset are nonapeptides (relatively small in molecular weight), we applied the same preprocessing pipeline as used for PLIs to enhance model representation accuracy.

(6) PNI Evaluation

The test set for evaluating PNI scoring capability under the few-shot scenario was constructed by randomly splitting the preprocessed PDBbind PNI data as described above.

(7) NLI Evaluation

The dataset for evaluating NLI scoring capability under the zero-shot scenario was derived from the preprocessed PDBbind NL data and further refined by identifying the RNA–ligand interaction subset, as defined in our previous work.

(8) Protein–Cyclic Peptide Test Set

Constructed based on the protein–cyclic peptide benchmark dataset CPSet proposed in our previous work[66], retaining only those complexes with explicit binding affinity annotations.

(9) Protein–Non-Peptidic Macrocycle Test Set

Constructed from PDBbind PLI by filtering complexes whose ligands are macrocyclic molecules (defined as containing at least one ring with $\geq$12 atoms, based on SMILES) and have explicit binding affinity annotations. All cyclic peptide ligands were manually excluded through curation.

(10) Protein–Carbohydrate Test Set

Constructed by using the protein–carbohydrate dataset curated by Nguyen et al. as the training set[71], with 43 entries from PCAPRED serving as an independent test set[72].

All test sets underwent the same filtering and interaction region extraction procedures as described for the pretraining and fine-tuning datasets. Complexes that could not be effectively processed were discarded. Detailed information on each test set is provided in Table S2. We used the model weights pretrained on PDBbind PLI+PPI data as the foundation for testing under various scenarios. For all docking and screening evaluations, no additional fine-tuning was performed; the pretrained weights were directly applied. For all scoring and ranking evaluations, task-specific fine-tuning was conducted using the corresponding datasets: protein–small molecule scoring/ranking, protein–cyclic peptide and protein-non-peptidic macrocycle evaluations were fine-tuned on PDBbind PLI; protein–protein scoring evaluations were fine-tuned on PDBbind PPI; and antigen–antibody and protein–carbohydrate evaluations were fine-tuned on their respective pre-split datasets. Notably, for nucleic acid–related tasks such as PNI and NLI, we included a small portion of PDBbind PNI data in the pretraining stage and subsequently fine-tuned on PDBbind PNI. This strategy ensured that the model could learn interaction patterns relevant to nucleic acid complexes during training.

## BioScore: a theoretical introduction

Many deep learning–based scoring functions utilize custom-designed descriptors to capture key ligand–target interactions, which are then input into prediction algorithms. However, such approaches may essentially overfit the mapping between input data and labels without achieving true generalization and typically lack docking and screening capabilities. In contrast, integrating AI techniques into traditional scoring paradigms—such as force field–based or statistical potential–based methods—aligns more closely with intuitive modeling principles. Unlike force field–based methods, which decompose binding free energy into multiple energy terms, statistical potential–based scoring functions calculate binding free energy by summing the statistical potentials of molecular pairs within the complex.

The earliest statistical potential approaches derived pairwise distance probability distributions by analyzing large datasets of protein–small molecule complex structures, identifying the frequency at which specific (key) atom pairs—each atom coming from the protein or the small molecule—appear at particular distances[25,29]. The underlying rationale is that if a specific atom pair is observed at a particular distance significantly more frequently in the experimentally curated complex structures than expected by random chance, this pair is likely to form a favorable interaction. In other words, the occurrence frequency of specific atom pair distances is assumed to correlate positively with their contribution to the binding phenomenon. This intuition can be further formalized mathematically. For continuous systems, the Boltzmann distribution can be expressed as:

$$p(x) = \frac{e^{\frac{-\varepsilon(x)}{KT}}}{Z} \tag{1}$$

Here, x represents the state of the biomolecular complex, such as the atomic coordinates, that can be used to describe the energy of the system as $\epsilon(x)$. When the partition function $Z$ is explicitly specified, p(x) is a probability density function, satisfying:

$$\int_{-\infty}^{\infty} p(x)dx = 1 \tag{2}$$

From the Boltzmann distribution above, one can then define the potential of mean force (PMF), such as $\varepsilon_{pmf}(y)$, where $y$ is a subset of degrees of freedom of interest—eg. Pairwise distance between atom pairs in our context—that can be extracted from the full molecular configuraton $x$. The PMF is related to the probability density via:

$$\varepsilon_{pmf}(y) = -KT\log \int p(x)\delta(m(x) - y)dx + C \tag{3}$$

Where $K$ is the Boltzmann constant, $T$ is the temperature, $m(x) = y$ being a function to extract the subset of degrees of freedom of interest, and $C$ is a normalization constant. This formulation establishes a direct relationship between distance distributions and PMF or interaction energy (loosely speaking). In practice, we do not rigorously calculate PMF by rigorously performing the integration in Eq. (3) in the context of building a statistical potential. Rather, we simply try to use the limited set of available structural data to invert the Boltzmann distribution and derive a PMF-like free energy contribution. If we can accurately determine or fit the distance distributions for different atom pairs, we can compute their corresponding statistical potential.

Deep learning techniques are particularly well-suited for modeling such distance distributions. By leveraging mixture density network, it is possible to fit distance distributions as a weighted sum of Gaussian components, providing a more flexible and accurate representation. This insight has led to the development of methods such as DeepDock[80], RTMScore[38], and GenScore[55]. While these earlier works were primarily limited to protein–small molecule systems due to data availability constraints, our study demonstrates that this inverse Boltzmann–inspired approach can also be extended to protein–protein, protein–nucleic acid, and other biomolecular complexes, highlighting its broad generalizability and value.

For deep learning–based scoring functions, how to effectively represent atom pair features is a critical challenge. Previous studies have shown that through careful feature selection, it is possible to achieve effective scoring of protein–small molecule complexes at a non-all-atom, fine-grained level. However, to generalize such approaches to arbitrary biomolecular complexes, using domain-specific encoding strategies for different biomolecules lacks generalizability.

To address this problem, there are two feasible approaches. The first is to represent all biomolecular components uniformly at the atomic level. Since atoms are the fundamental building blocks of all biomolecules, this approach is intuitive. However, atomic-level encoding also has significant limitations: biomolecules differ greatly in their molecular forms, and "block"-level units—such as amino acids in proteins or nucleotides in nucleic acids—encode much richer, domain-specific information than atoms alone. This is why biomolecular language models such as ESM[81] and DNABERT[82] adopt block-level units as their fundamental sequence representations; their success highlights the importance of such heuristics in molecular representation learning.

The second approach is to exclusively encode molecules at the block level, for instance, representing proteins only at the residue level. This strategy can work well within specific biomolecular systems. For example, RTMScore[38] employs amino acids as the basic encoding unit for proteins to reduce computational complexity. However, the high specificity of block-level representations limits the transferability of information across different domains.

Our goal is to develop a foundational scoring function that not only supports arbitrary biomolecular inputs but also enables cross-system data complementarity—an essential requirement to address the practical challenge of limited data availability beyond protein–small molecule systems.

In summary, while encoding strategies based solely on atoms or blocks are feasible, they are not optimal for building a general-purpose scoring function. To overcome these limitations, we propose a cross-domain molecular representation that integrates dual-scale atomic and block-level information. Specifically, we decompose any molecular entity into block units: for small molecules, blocks correspond to atoms; for proteins, blocks correspond to amino acid residues; and for nucleic acids, blocks correspond to nucleobases. The entire molecular complex is then represented as a unified graph $\mathcal{G} = (\mathcal{V}, \mathcal{E})$, where the node set $\mathcal{V} = \{(H_i, \vec{X_i}) | 1 \leq i \leq B\}$ corresponds to the block units. Each node contains a feature matrix and a coordinate matrix. For a block consisting of $n_i$ atoms, the feature matrix $H_i \in \mathbb{R}^{n_i \times d_h}$ represents the feature vectors of the atoms, and the coordinate matrix $X_i \in \mathbb{R}^{n_i \times 3}$ encodes the corresponding 3D atomic coordinates. The feature vector for each atom is constructed by combining three learnable embeddings: the atomic type, the block type it belongs to, and the relative position of the atom within the block. Formally, we define:

$$d(i,j) = min \left\{ \left\| \vec{X_i}[p] - \vec{X_j}[q] \right\|_2 \bigg| 1 \leq p \leq n_i,\ 1 \leq q \leq n_j \right\} \quad (4)$$

It is important to emphasize that the key difference between our approach and previous general-purpose biomolecular complex representation learning methods lies in our introduction of a interface-masking encoding strategy. Specifically, we represent the entire complex as a single graph but construct edges only within individual molecular entities, deliberately excluding inter-molecular edges. Prior works on general biomolecular complex representation learning were primarily designed to better represent different biomolecular systems and interaction interfaces, without a specific focus on the unique requirements of scoring function design for complexes.

As previously discussed, in the design of statistical potential–based scoring functions, appropriate representation learning should aim to facilitate the neural network—specifically, the mixture density network—in learning a reasonable mapping between atom pair representations and the probability density distributions of their distances. Our study reveals that in such statistical potential–based frameworks, explicitly exposing the inter-molecular edge information during atom pair representation learning leads to inferior mapping results.

This contrasts with the conventional wisdom in prior general biomolecular complex representation studies, which consistently emphasize the importance of inter-molecular edge information for representing biomolecular interactions. While this design choice may seem intuitive (as interaction edges appear essential for encoding biomolecular interfaces), our investigation demonstrates that for statistical potential–based scoring, incorporating interaction edge information actually bears a detrimental effect.

Further analysis shows that within the statistical potential modeling framework, inter-molecular edges, though commonly used in representation learning to strengthen the modeling of atom pair relationships, introduce significant modeling biases. These explicit interaction priors cause the model to receive strong, structure-specific signals about the native conformation at the input stage, leading to overly concentrated predictions around the true distances in the training samples. This results in a loss of smoothness and generalization in the learned distance distributions, effectively causing a structural overfitting. We support this analysis with evidence from the validation loss curves (Figure S1). Moreover, representing the entire complex as a single graph—rather than adopting a dual-branch architecture where each molecular entity is modeled independently—offers a key advantage: it provides a unified representation format for different types of complexes. Without this design, each biomolecular system would require a dedicated encoder, undermining the model's generality and limiting its capacity to integrate information across domains.

In addition to the interface-masking strategy, another noteworthy technical detail of our approach is the design of a distance-threshold-based edge construction strategy for both intra- and inter-molecular edges. Traditional edge construction strategies generally fall into two categories. The first is the exhaustive enumeration approach, where all pairwise distances between atoms in interacting biomolecules are precomputed. This method was employed in our previous works, RTMScore[38,55] and GenScore. However, this approach suffers from low computational efficiency, particularly when extending beyond protein–small molecule complexes (which typically feature binding interfaces with median sizes of 300–1500 Å²) to larger biomolecular systems[16], such as protein–protein complexes, where the median interface size ranges from 2000 to 4000 Å². The second common strategy is the k-nearest neighbors approach, where a fixed number of intra- and inter-molecular edges are assigned to each atom. While more efficient, this method discards valuable information inherent in the number of edges itself. For example, in classical statistical potential–based methods, the total binding free energy is computed by summing the contributions from key atom pairs, implicitly capturing the principle that a larger number of critical interaction edges tends to correlate with stronger binding affinity.

To preserve this information and enable the incorporation of interaction edge count–aware confidence scores, we predefined distance thresholds for edge construction: for inter-molecular edges, a uniform threshold of 8 Å was applied across all biomolecular complex systems, such that only atom pairs within 8 Å (from different molecular entities) are connected by inter-molecular edges. For intra-molecular edges, we applied system-specific thresholds: 2 Å for small molecules and 10 Å for proteins and nucleic acids. Compared to the KNN approach, our distance-threshold strategy preserves meaningful variations in edge counts. Compared to the exhaustive enumeration approach, it significantly improves computational efficiency while maintaining the critical interaction patterns required for accurate scoring.

To enable a unified representation of such complex systems, we adopted GET (Geometric Equivariant Transformer) as the encoder, following the design introduced by Kong et al.[41] This framework effectively captures both domain-specific hierarchical structures and general cross-domain physical interaction properties. GET is composed of equivariant dual-level attention modules, feedforward modules, and layer normalization modules, with each component being E(3)-equivariant. In each layer of GET, the equivariant dual-level attention module first jointly models block-level and atom-level interactions. The equivariant feedforward module then updates atomic features within each block by incorporating geometric information from the block as a whole. Finally, the equivariant layer normalization module ensures invariance of the model under molecular rotations and translations. Specifically, in the equivariant dual-level attention module, for two blocks $i$ and $j$ composed of $n_i$ and $n_j$ atoms, respectively, the atom-level cross-attention between blocks $i$ and $j$ is computed as:

$$R_{ij}[p,q] = MLP_A\left(Q_i[p], K_j[q], RBF(D_{ij}[p,q], e_{ij})\right) \tag{5}$$

$$\alpha_{ij} = Softmax(R_{ij}W_A) \tag{6}$$

Here, $R_{ij}[p,q]$ represents the interaction information matrix between each pair of atoms from blocks $i$ and $j$. $MLP_A$ is used to compress the input features into interaction information vectors. $Q_i[p]$ denotes the query feature of atom $p$ in block $i$, $K_j[q]$ denotes the key feature of atom $q$ in block $j$, $D_{ij}[p,q]$ is the Euclidean distance between atoms $p$ and $q$, and $e_{ij}$ represents the distance feature between blocks $i$ and $j$. The pairwise atomic distances are embedded using radial basis functions (RBF). A learnable linear projection weight $W_A \in \mathbb{R}^{d_r \times 1}$ is applied to map $R_{ij} \in \mathbb{R}^{n_i \times n_j \times d_r}$ into scalar values. These scalar values are then passed through a softmax function to obtain the atom-level cross-attention $\alpha_{ij} \in \mathbb{R}^{n_i \times n_j}$ between blocks $i$ and $j$. The block-level cross-attention between blocks $i$ and $j$ is computed by aggregating the atom-level attention values as follows:

$$r_{ij} = \frac{1}{n_i n_j} \sum_{p=1}^{n_i} \sum_{q=1}^{n_j} R_{ij}[p,q] \tag{7}$$

$$\beta_{ij} = \frac{exp(r_{ij}W_B)}{\sum_{k \in \mathcal{N}(i)} exp(r_{ik}W_B)} \tag{8}$$

The subsequent step updates the hidden states and coordinates of each atom within a block,

$$H'_i[p] = H_i[p] + \sum_{j \in \mathcal{N}(i)} \beta_{ij} MLP_H(m_{ij,p}) \tag{9}$$

$$\vec{X}'_i[p] = \vec{X}_i[p] + \sum_{j \in \mathcal{N}(i)} \beta_{ij}(MLP_X(m_{ij,p}) \cdot \vec{m}_{ij,p}) \tag{10}$$

Taking block $i$ as an example, the hidden state of each atom $p$ is updated by computing a weighted sum of the original feature $H_i[p]$ and the aggregated information $m_{ij,p}$ from its neighboring blocks $j$ where the weights for different neighboring blocks are determined by the block-level attention coefficient $\beta_{ij}$. For coordinate updates, the direction shift vector $\vec{m}_{ij,p}$ is scaled by an element-wise product with the scalar coefficient $MLP_X(m_{ij,p})$, and the resulting vectors are aggregated across neighboring blocks using the same block-level attention weights $\beta_{ij}$. In the subsequent equivariant feedforward network module, the hidden states and coordinates of each atom within a block are further updated, enabling each atom to learn the geometric information of its corresponding block,

$$h' = h + MLP_h(h, h_c, RBF(\|(\vec{x} - \vec{x_c})\|_2)) \tag{11}$$

$$\vec{x}' = \vec{x} + (\vec{x} - \vec{x_c})MLP_x(h, h_c, RBF(\|(\vec{x} - \vec{x_c})\|_2)) \tag{12}$$

Here, $h_c$ and $\vec{x_c}$ denote the hidden state and coordinates of the centroid of the block to which the atom belongs. To ensure E(3)-equivariance, we apply equivariant layer normalization to the atomic coordinates. Specifically, we first extract the centroid of the entire graph, $\mathbb{E}[\vec{X}]$, where $\vec{X}$ represents the coordinates of all atoms across all blocks in the graph. We then apply layer normalization to both the hidden states and the coordinates as follows:

$$h' = \frac{h - \mathbb{E}[h]}{\sqrt{Var[h]}} \cdot \gamma + \beta \tag{13}$$

$$\vec{X}' = \frac{\vec{X} - \mathbb{E}[\vec{X}]}{\sqrt{Var[\vec{X} - \mathbb{E}[\vec{X}]]}} \cdot \sigma + \mathbb{E}[\vec{X}] \tag{14}$$

where $\gamma, \beta, \sigma$ are learnable parameters, and $Var[\cdot]$ denotes the variance computed relative to the centroid. This process ensures that, after subtracting the centroids of all atoms, the coordinates are normalized and scaled according to a standard Gaussian distribution.

Based on the above framework, we can effectively use a mixture density network to model the mapping between block pair representations and their corresponding distance probability density distributions. Specifically, we first concatenate the features of block pairs within the complex (denoted as receptor and ligand to represent the two molecular entities in the complex):

$$h_{i,j}^{complex} = Concat(h_i^{ligand}, h_j^{recptor}) \tag{15}$$

The concatenated block pair feature is then transformed via an Multilayer Perceptron (MLP):

$$I_{i,j}^{complex} = MLP(h_{i,j}^{complex}) \tag{16}$$

The parameters of the Gaussian components in the mixture density network—namely the means $\mu$, standard deviations $\sigma$, and mixture coefficients $\alpha$—are obtained via three simple linear layers applied to $I_{i,j}^{complex}$:

$$\mu_{i,j} = ELU\left(W_\mu I_{i,j}^{complex} + b_\mu\right) + 1.0 \tag{17}$$

$$\sigma_{i,j} = ELU\left(W_\sigma I_{i,j}^{complex} + b_\sigma\right) + 1.1 \tag{18}$$

$$\alpha_{i,j} = Softmax\left(W_\alpha I_{i,j}^{complex} + b_\alpha\right) \tag{19}$$

where $W_\mu$, $W_\sigma$, $W_\alpha$, $b_\mu$, $b_\sigma$, $b_\alpha$ are learnable parameters of the respective linear layers. The probability density distribution of the distance for a single block pair can then be expressed as:

$$P\left(d_{i,j}^{complex}\middle|I_{i,j}^{complex}\right) = \sum_{n=1}^{10} \alpha_{i,j,n} \mathcal{N}\left(d_{i,j}^{complex}\middle|\mu_{i,j,n}, \sigma_{i,j,n}\right) \tag{20}$$

As previously shown in Equation (3), based on the inverse Boltzmann relationship between probability density and energy, the calculation of statistical-potential-approximated binding free energy in prior studies typically follows the formula:

$$\Delta G = -KT \sum_{ij}^{d_{ij} < cutoff} \ln\left(P(d_{ij}|I_{ij})\right) \tag{21}$$

where the cutoff indicates that only block pairs within a predefined distance threshold contribute to the final binding free energy calculation. However, based on the fundamental equation for binding free energy, we know that:

$$\Delta G = G_{bound} - G_{unbound} = E_{bound}^{inter} + E_{bound}^{intra} - E_{unbound}^{intra} \tag{22}$$

Traditional statistical-potential–based scoring functions essentially compute only the interaction term under the bound state:

$$E_{bound}^{inter} = -KT \sum_{i,j}^{d_{i,j}^{complex} < cutoff} \ln\left(P\left(d_{i,j}^{complex}\middle|I_{i,j}^{complex}\right)\right) \tag{23}$$

In other words, they estimate the relative binding free energy by ignoring the contribution arising from the conformational rearrangements of individual molecules upon complex formation. To approximate the absolute binding free energy, these methods assume no significant energetic differences due to conformational changes in the unbound state, i.e. many methods simply ignore $E_{bound}^{intra} - E_{unbound}^{intra}$ in Eq. (22). To address this limitation, we propose a dual-tower scoring module that separately handles docking/screening and scoring/ranking tasks. First, we retain the core principle of the traditional statistical potential approach as summarized above. However, we explicitly treat the computed energy as a relative binding free energy rather than an absolute one. At the same time, we recognize an inherent drawback of the traditional statistical potential methods: they compute the total binding energy by summing the log-transformed probability densities of key atom pairs (i.e., those within a predefined distance cutoff). This design can inadvertently assign artificially high scores to implausible complex conformations, where each individual edge may have a low score, but the accumulation of many unreasonable edges leads to an overall unreasonably high score. To mitigate this issue, we simply weight the summation by the number of contributions. This change prevents the model from overestimating the contributions of unfavorable conformations due to sheer amount of atom pairs. Meanwhile, we preserve a key advantage of the summation formulation—namely, the implicit positive correlation between the number of effective interaction edges and binding affinity. To capture this, we introduce an interaction edge count–aware confidence score as an additional scoring term. Specifically:

$$U_{cplx} = \frac{1}{n_{i,j|d_{i,j}^{complex}<cutoff}} \sum_{ij}^{d_{i,j}^{complex}<cutoff} ln\left(P\left(d_{i,j}^{complex}\middle|I_{i,j}^{complex}\right)\right) \tag{24}$$

$$Score = -U_{cplx} + log\left(n_{i,j|d_{i,j}^{complex}<cutoff}\right) \tag{25}$$

We demonstrate that the modified output formulation significantly improves docking and screening capabilities in both PPI and PLI tasks, exhibiting strong generalizability across different biomolecular complex systems. We designate this output score as the first output mode of the scoring function, specifically tailored for docking and screening workflows across arbitrary biomolecular complexes. This score provides the capability to (1) distinguish native-like from non-native binding conformations (docking) and (2) differentiate active from inactive ligands (screening) across various biomolecular systems. Furthermore, we extend the block pair representations learned by the model to define a second output mode for the scoring function. Here, an MLP is used to learn the mapping between block pair representations and binding free energy values. The final output score is computed as the mean of these pairwise contributions, while the interaction edge count–aware confidence score is incorporated as an auxiliary output term. This second output mode is used for scoring and ranking workflows across arbitrary biomolecular complexes:

$$U_{cplx} = \frac{1}{n_{i,j|d_{i,j}^{complex}\leq cutoff}} \sum_{i,j}^{d_{i,j}^{complex}\leq cutoff} MLP\left(I_{i,j}^{complex}\right) \tag{26}$$

$$Score = -U_{cplx} + \alpha_\theta * log\left(n_{i,j|d_{i,j}^{complex}<cutoff}\right) \tag{27}$$

where $\alpha_\theta$ are learnable parameters.

To achieve a balance among the four capabilities of scoring, ranking, docking, and screening, we propose a pretraining–fine-tuning strategy. In the pretraining stage, we utilize only the structural data from various biomolecular complexes to train the mixture density network module. By minimizing the negative log-likelihood loss to maximize the probability density values of block pair distances, the model learns generalizable structural features that enable docking and screening capabilities. The loss function for this stage is defined as:

$$\mathcal{L}_{Pretrain} = \mathcal{L}_{MDN} = \frac{1}{n_{i,j|d_{i,j}^{complex}\leq cutoff}} \sum_{i,j}^{d_{i,j}^{complex}\leq cutoff} -lnP\left(d_{i,j}^{complex}\middle|I_{i,j}^{complex}\right) \tag{28}$$

In the subsequent fine-tuning stage, labeled data from different biomolecular complexes (i.e., complexes with both structural and binding affinity annotations) are used for task-specific fine-tuning within each biomolecular system. The loss function for this stage is defined as:

$$\mathcal{L}_{Finetune} = a \times \mathcal{L}_{MSE} + b \times \mathcal{L}_{Correlation} + c \times \mathcal{L}_{MDN} \tag{29}$$

where $a$, $b$, $c$ are empirically defined hyperparameters, with default values set to 0.5, 5, and 1, respectively. During pretraining, the distance threshold is set to 7 Å, whereas in the fine-tuning stage and the dual-tower scoring outputs, it is set to 5 Å. Experiments were conducted on NVIDIA L20 GPUs (48 GB VRAM). The model is trained using the Adam optimizer. The hyperparameter settings for both the pretraining and fine-tuning stages are summarized in Table S18.

# Acknowledgement

CYH acknowledges support from National Ke Research and Development Program of China (2024FA1306400) and Natural Science Foundation of China (22373085).

# Tables

Table 1. PPI Benchmark Test Results[a].

| Model | Scoring | | | Docking | | | | | | Screening | | | |
|---|---|---|---|---|---|---|---|---|---|---|---|---|---|
| | Pearson | RMSE | MAE | Spearman (All) | Spearman (Top 20) | SR (Top 1) | SR (Top 5) | SR (Top 10) | Hit (Top 1) | $SR_{1\%}$ | $SR_{5\%}$ | $SR_{10\%}$ | Average Rank |
| GET | 0.25 | 3.94 | 3.37 | 0.40 | -0.06 | 0.00% | 0.00% | 3.80% | 26.58% | 1.27% | 5.06% | 12.66% | 35.57 |
| GNN-DOVE | -0.27 | 7.64 | 7.34 | 0.11 | -0.15 | 0.00% | 1.27% | 2.53% | 24.05% | 2.53% | 6.33% | 15.19% | 36.34 |
| DProQA | 0.01 | 7.60 | 7.37 | 0.16 | 0.17 | 5.71% | 15.71% | 20.00% | 35.71% | 6.06% | 9.09% | 25.76% | 30.21 |
| VoroMQA | -0.10 | 7.48 | 7.17 | **0.74** | 0.28 | 24.05% | 41.77% | 63.29% | 69.62% | 1.27% | 6.33% | 16.46% | 26.22 |
| ZRANK2 | -0.08 | 131.75 | 97.39 | 0.30 | -0.18 | 0.00% | 0.00% | 0.00% | 26.58% | 5.06% | 8.86% | 21.52% | 31.13 |
| MINT | **0.49** | **1.82** | **1.32** | - | - | - | - | - | - | 2.53% | 6.33% | 16.46% | 36.15 |
| BioScore | 0.46 | 2.24 | 1.65 | 0.64 | **0.55** | **100.00%** | **100.00%** | **100.00%** | **100.00%** | **15.19%** | **32.91%** | **41.77%** | **21.54** |

[a]Both GET and MINT were trained using the same training data as BioScore. Since MINT is designed exclusively for protein–protein complexes, it was fine-tuned on the same PPI training data as BioScore, using the pretrained weights as released in the original MINT publication. In contrast, GNN-DOVE and DProQA were evaluated using the original model weights provided in their respective publications, as their training strategies differ from ours. Additionally, MINT accepts only sequence-based inputs and therefore lacks the capability to score different conformations of the same sequence, which limits its docking evaluation capability.

# Figures

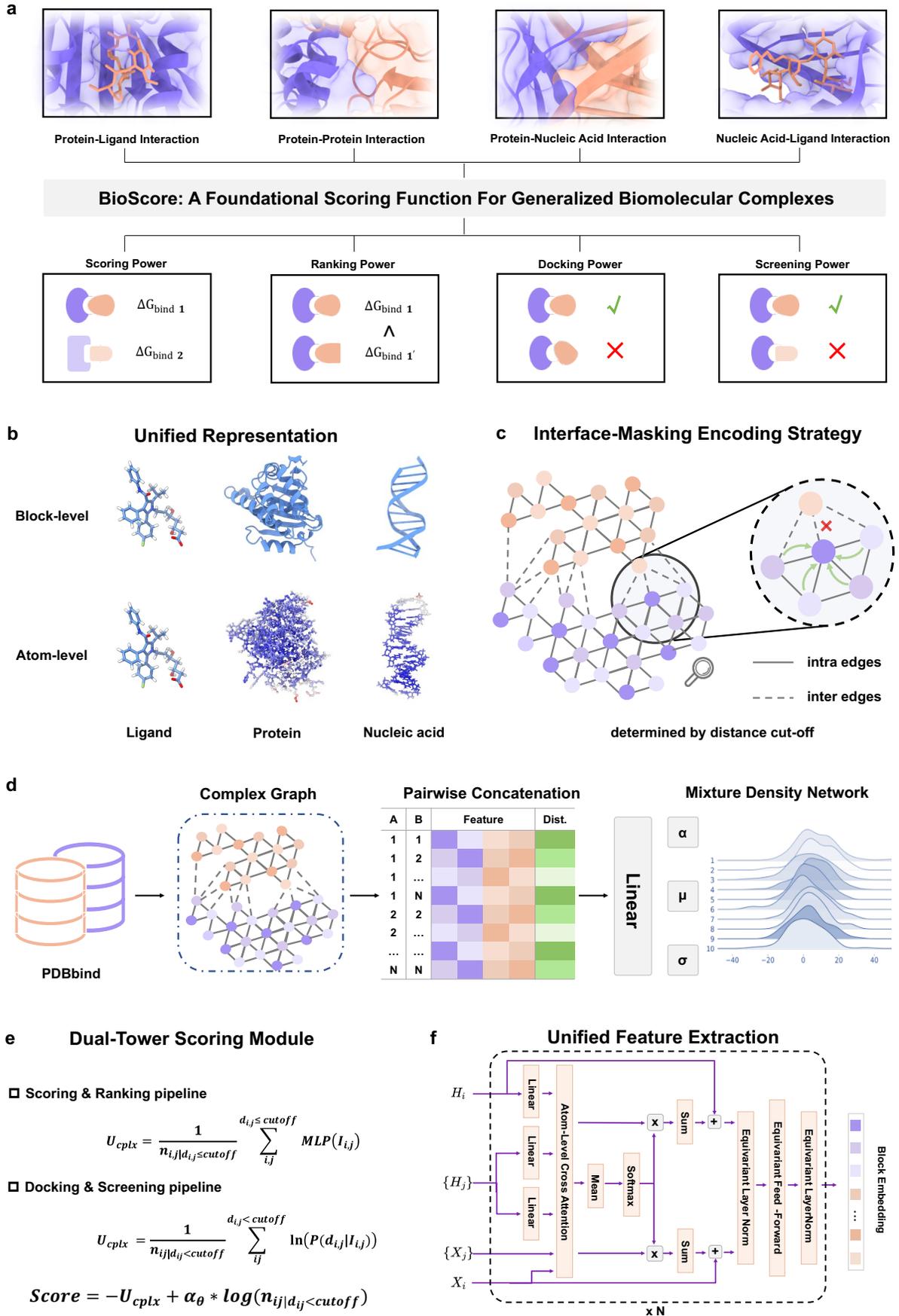

**Figure 1. Overview of BioScore.**

(A) BioScore Description. BioScore accepts complex structure inputs across diverse biomolecular systems and provides four core capabilities: scoring, ranking, docking, and screening.

(B) Dual-Scale Representation. BioScore incorporates both atomic- and block-level information, where blocks correspond to atoms for small molecules, amino acid residues for proteins, and nucleotides for nucleic acids.

(C) Interface-Masking Encoding and Distance-Threshold-Based Edge Construction. BioScore intentionally masks inter-molecular edges and retains only intra-molecular edges, enhancing the model's ability to capture and fit true spatial distances.

(D) Pretraining Strategy. The model trains a mixture density network by minimizing the negative log-likelihood to maximize the probability density of atom pair distances, enabling docking and screening capabilities.

(E) Fine-Tuning Strategy and Dual-Tower Scoring Module. BioScore employs a dual-tower design, with one branch for docking and screening tasks and another for scoring and ranking tasks. In the docking/screening pathway, the model constructs statistical potentials based on the inverse Boltzmann distribution, outputs energy via averaging (instead of summation), and incorporates an interaction edge count–aware confidence score as an auxiliary term. In the scoring/ranking pathway, the model fits the nonlinear relationship between atom pair representations and binding free energy using a neural network, outputs the mean score as the primary output, and adds an interaction-aware confidence term.

(F) Feature Extraction Module. BioScore integrates a general equivariant Transformer for feature extraction.

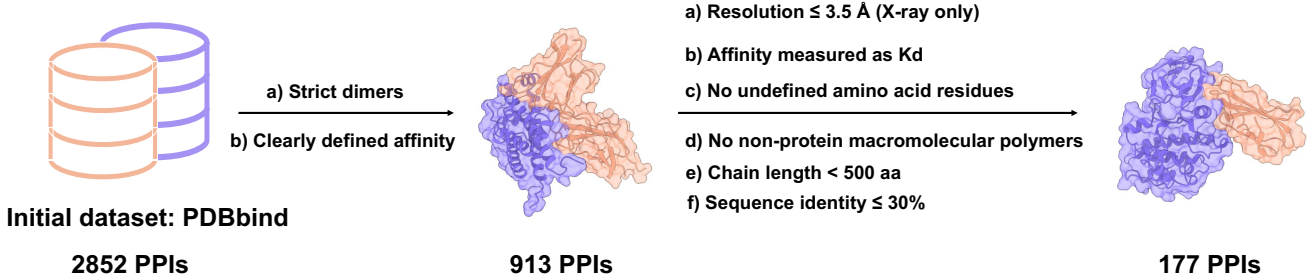
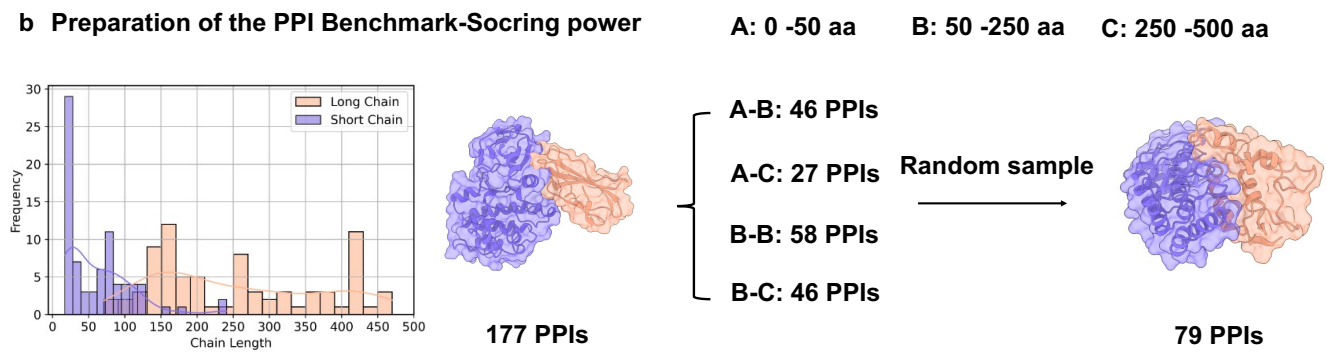
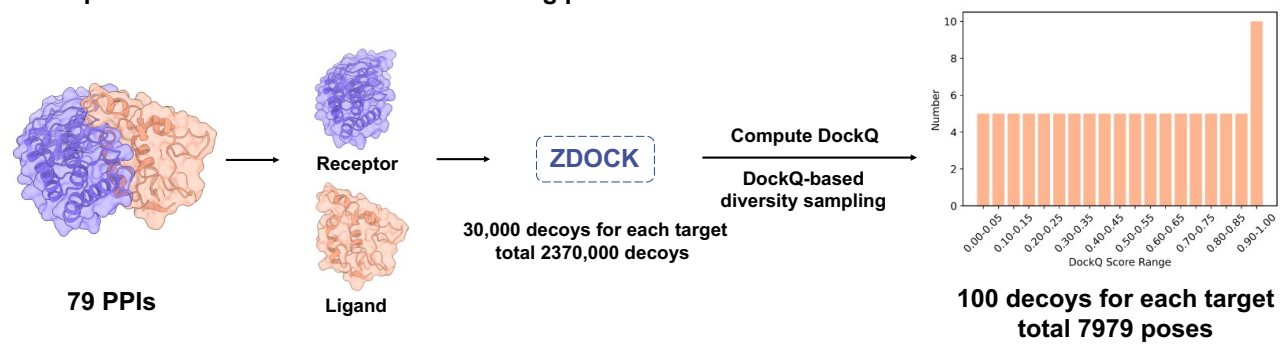
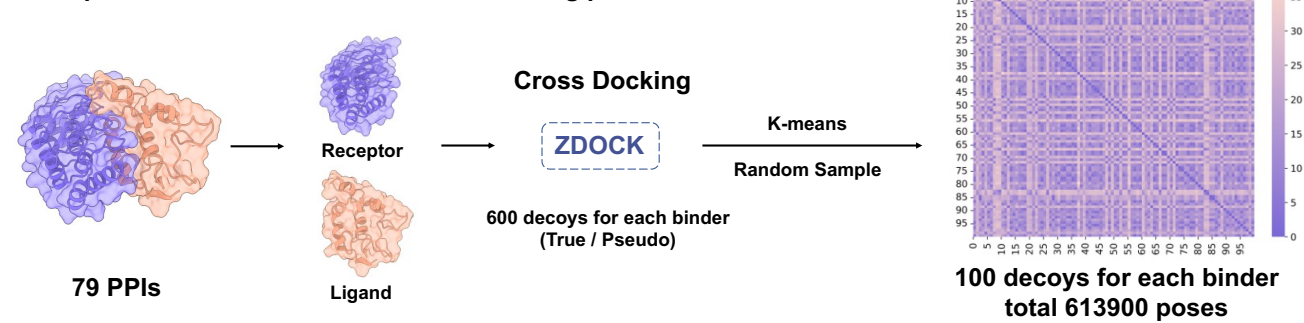

**Figure 2. Workflow for Constructing the PPI Benchmark.**
(A) Filtering and selection of the foundational complex dataset. A total of 177 protein–protein complexes were selected from PDBbind based on multiple criteria, and further filtered by chain length diversity to obtain a final set of 79 complexes as the foundational dataset.

(B) Construction of the scoring test set. The scoring test set consists of 79 native complexes with corresponding experimental binding affinity annotations.

(C) Construction of the docking test set. For each native complex, ZDOCK was used to generate decoy conformations, and a total of 7,979 decoys were selected to ensure a balanced distribution of DockQ scores.

(D) Construction of the screening test set. The screening test set was created by cross-docking 79 receptor–ligand pairs. K-means clustering and sampling were used to generate a total of 613,900 conformations.

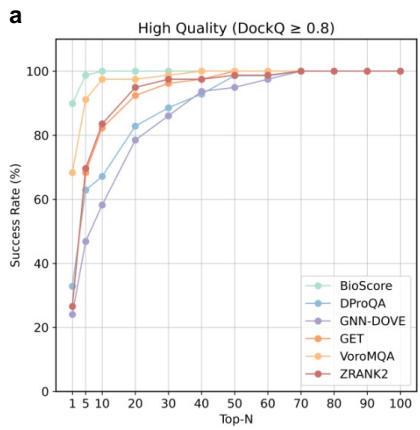
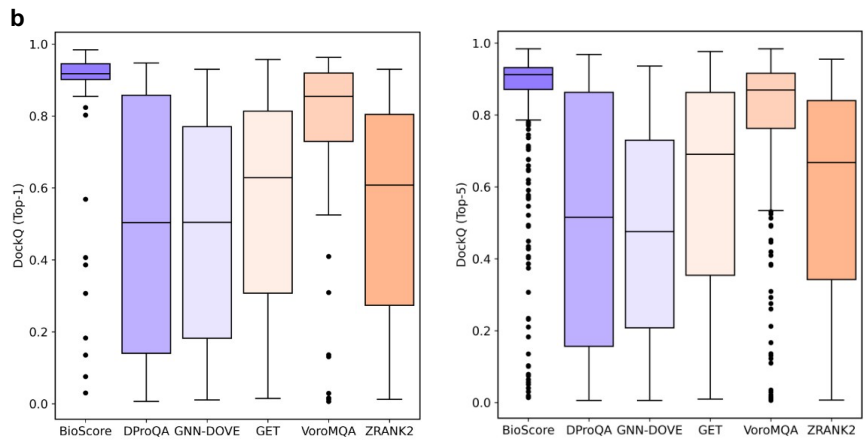
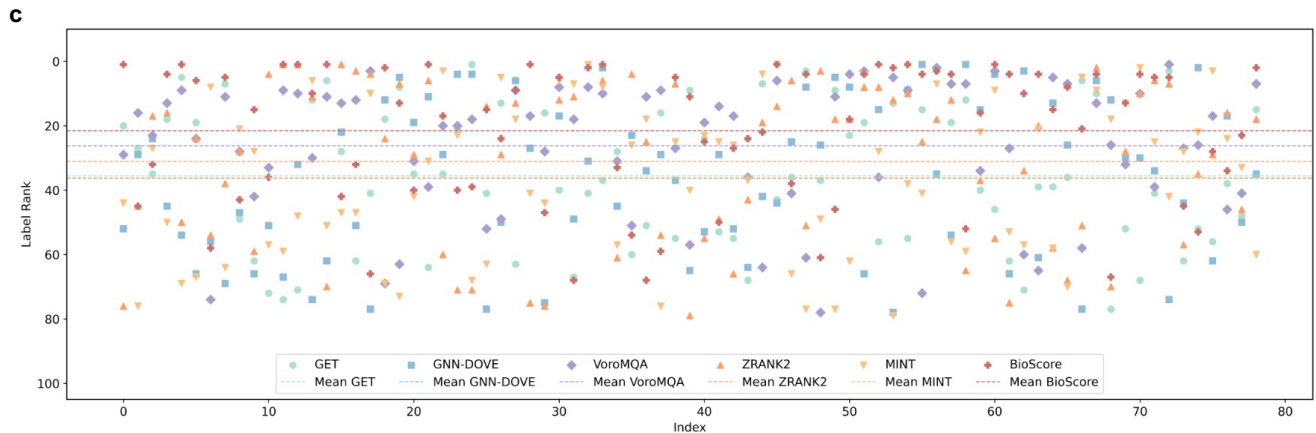
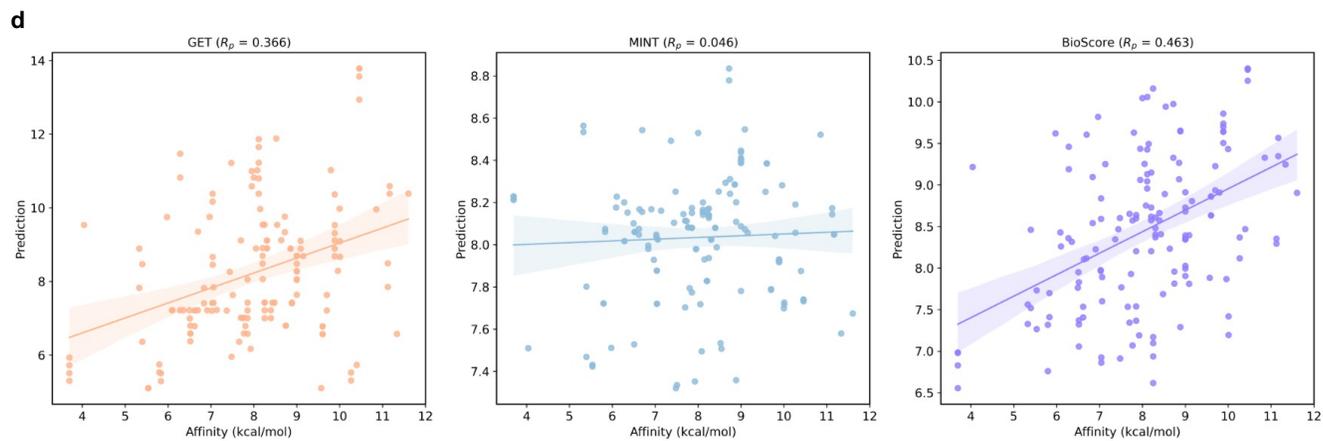
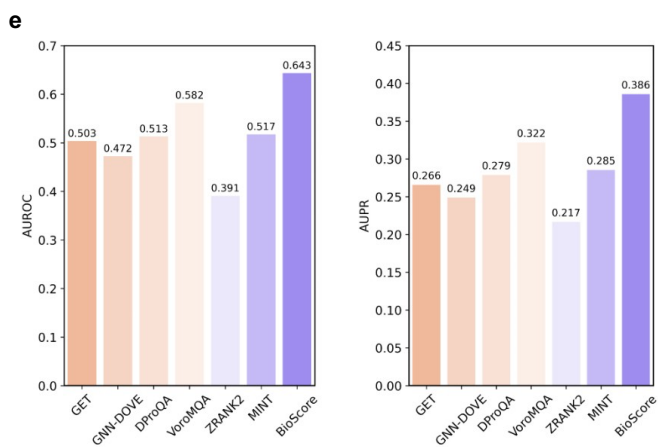
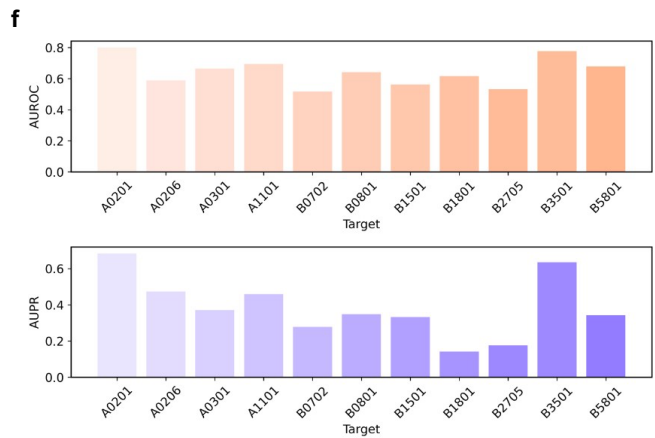

**Figure 3. Comprehensive Evaluation of BioScore on Protein–Protein Complex Systems.**

(A) Docking performance on the PPI Benchmark. Success rate of identifying high-quality conformations (DockQ score ≥ 0.8) based on scoring.

(B) Docking performance on the PPI Benchmark. Distribution of conformation quality (DockQ scores) for top 1 and top 5 ranked decoys.

(C) Screening performance on the PPI Benchmark. Distribution of rank values for the true ligand proteins of 79 target proteins; lower rank values indicate better ranking.

(D) Antigen–antibody scoring performance. Binding affinity prediction results of BioScore and baseline models.

(E) Peptide–MHC-I screening evaluation. Test results of different models based on AUROC and AUPR metrics.

(F) BioScore performance on peptide–MHC-I screening. Detailed evaluation metrics for different targets.

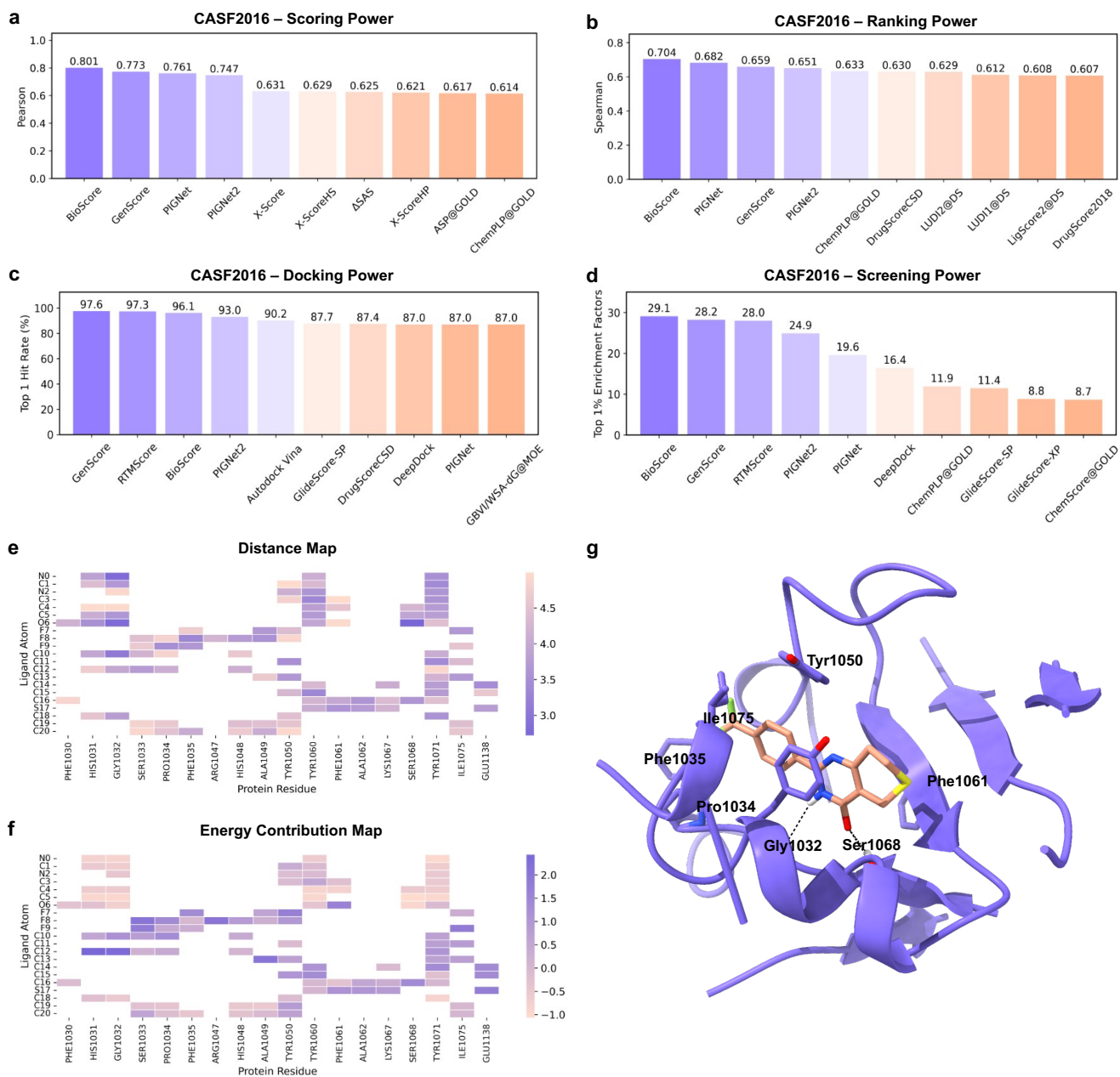

**Figure 4. BioScore Performance on the Protein–Small Molecule Benchmark CASF-2016.**

(A) Scoring performance on CASF-2016, evaluated by the Pearson correlation coefficient.

(B) Ranking performance on CASF-2016, evaluated by the Spearman correlation coefficient.

(C) Docking performance on CASF-2016, evaluated by the docking hit rate (including native poses).

(D) Screening performance on CASF-2016, evaluated by enrichment factors.

(E) The distance map of residue-atom interactions for the 3KR8 complex.

(F) The energy contribution map of residue-atom interactions for the 3KR8 complex, as calculated by BioScore.

(G) Visualization of key molecular interactions in the 3KR8 complex. Highlighted residues denote critical interaction sites; hydrogen bonds are represented by black dashed lines.

Only the top 10 methods for each test are shown here. Additional methods and results for all test metrics are provided in the Supplementary Information.

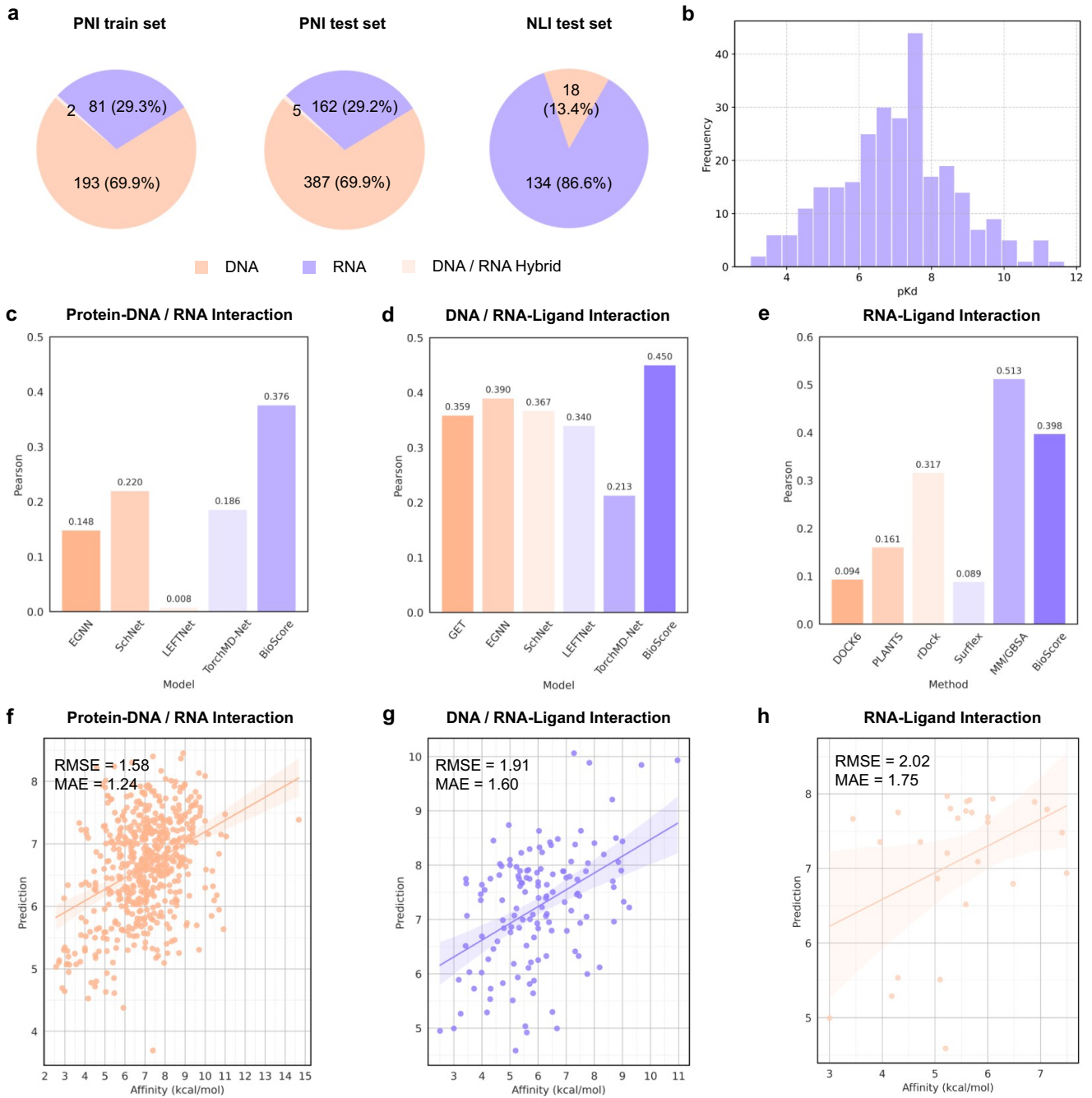

**Figure 5. BioScore Evaluation on Protein–Nucleic Acid and Nucleic Acid–Small Molecule Complexes.**

(A) Data distribution of DNA, RNA, and mixed DNA–RNA complexes across the PNI training set, PNI test set, and NLI test set.

(B) Distribution of experimental binding affinity values in the PNI training set.

(C) Scoring performance on the PNI test set, evaluated by the Spearman correlation coefficient.

(D) Scoring performance on the NLI test set, evaluated by the Pearson correlation coefficient.

(E) Scoring performance on the RNA–small molecule subset of the NLI test set, evaluated by the Pearson correlation coefficient.

(F) Detailed scoring results of BioScore on the PNI test set.

(G) Detailed scoring results of BioScore on the NLI test set.

(H) Detailed scoring results of BioScore on the RNA–small molecule test set.

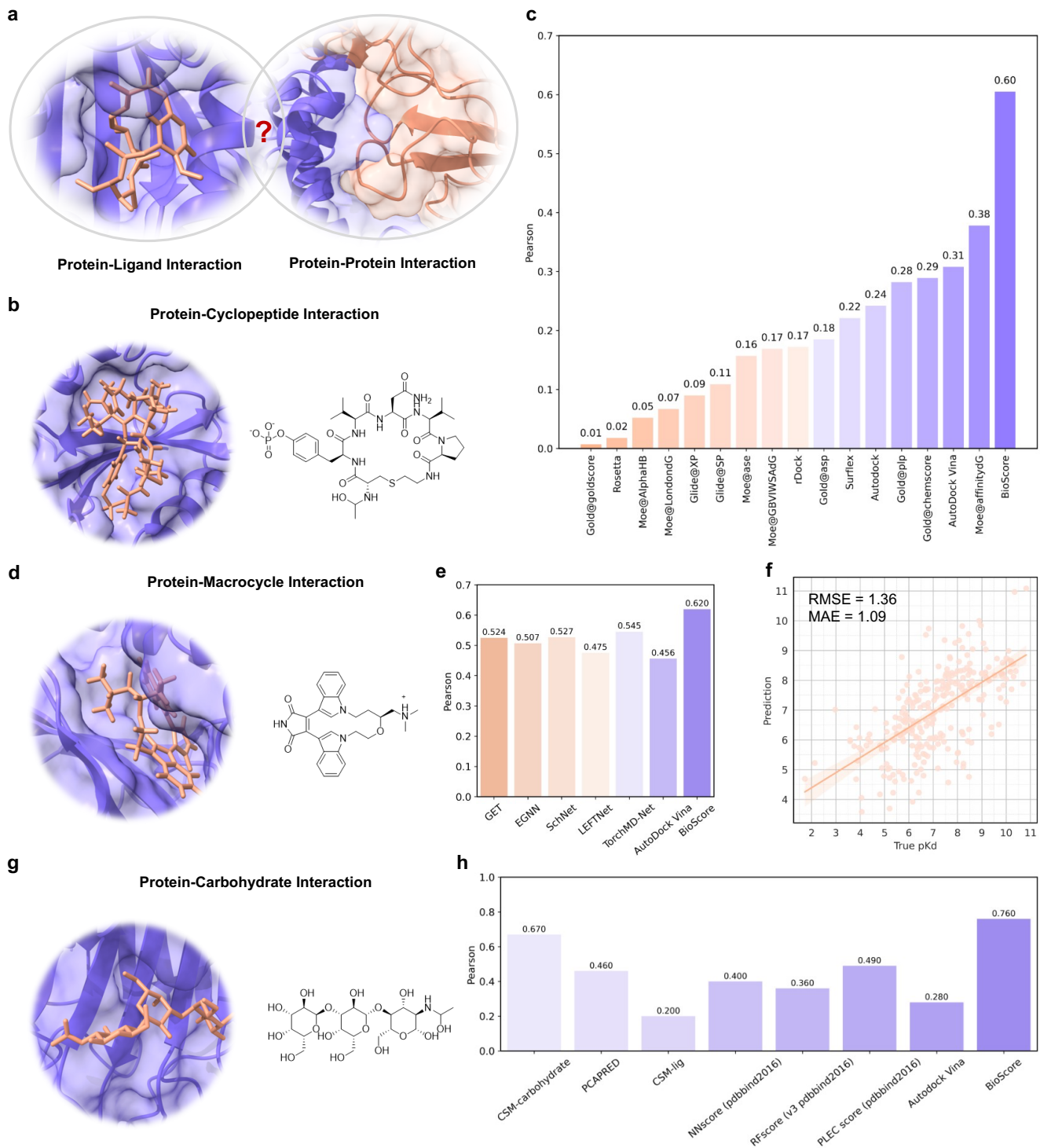

**Figure 6. BioScore Performance on Specialized Biomolecular Complexes (Protein–Cyclic Peptide, Protein–Non-Peptidic Macrocycle, and Protein–Carbohydrate Systems).**

(A) Schematic of the chemical space positions of cyclic peptides and non-peptidic macrocycles in complex with proteins, highlighting their location at the interface between protein–protein and protein–small molecule systems.

(B) Structural illustration of protein–cyclic peptide complexes and standalone cyclic peptide molecules.

(C) Protein–cyclic peptide scoring performance on the CPSet benchmark.

(D) Structural illustration of protein–non-peptidic macrocycle complexes and standalone non-peptidic macrocycle molecules.

(E) Scoring performance on the protein–non-peptidic macrocycle benchmark.

(F) Structural illustration of protein–carbohydrate complexes and standalone carbohydrate molecules.

(G) Scoring performance on the protein–carbohydrate benchmark.

# Supporting Information

# BioScore: A Foundational Scoring Function For Diverse Biomolecular Complexes

Yuchen Zhu[#,1], Jihong Chen[#,1], Yitong Li[#,1], Xiaomin Fang[2], Xianbin Ye[2], Jingzhou He[2], Xujun Zhang[1], Jingxuan Ge[1], Chao Shen[3], Xiaonan Zhang[2,*], Tingjun Hou[1,3,4,*], Chang-Yu Hsieh[1,3,4,*]

[1] College of Pharmaceutical Sciences, Zhejiang University, Hangzhou, 310058, China

[2] Baidu Online Network Technology (Beijing) Co., Ltd. Beijing, 100085, China

[3] The First Affiliated Hospital, College of Medicine, Zhejiang University, Hangzhou, 310058, China

[4] Zhejiang Provincial Key Laboratory for Intelligent Drug Discovery and Development, Jinhua, 321016, Zhejiang, China

Chang-Yu Hsieh

E-mail: kimhsieh@zju.edu.cn

Tingjun Hou

E-mail: tingjunhou@zju.edu.cn

Xiaonan Zhang

E-mail: zhangxiaonan@baidu.com

#These authors contributed equally.

# 1. Supplementary Tables

Table S1. Summary of Baseline Methods

| Method | Test Type |
|---|---|
| GET | PLI,PPI,PNI,NLI |
| GNN-DOVE | PPI |
| DProQA | PPI |
| VoroMQA | PPI |
| ZRANK2 | PPI |
| MINT | PPI |
| AutoDock Vina | PLI |
| RTMScore | PLI |
| EquiScore | PLI-DUD-E/DEKOIS2.0 |
| Glide SP | PLI |
| PIGNet | PLI |
| 3D-GNN | PLI-DUD-E/DEKOIS2.0 |
| TANKBind | PLI-DUD-E/DEKOIS2.0 |
| DeepDock | PLI |
| Kdeep | PLI-DUD-E/DEKOIS2.0 |
| deltaVinaRF | PLI |
| RFScorev4 | PLI-DUD-E/DEKOIS2.0 |
| NNScore2.0 | PLI-DUD-E/DEKOIS2.0 |
| OnionNet | PLI-DUD-E/DEKOIS2.0 |
| pafuncy | PLI-DUD-E/DEKOIS2.0 |
| PIGNet2 | PLI |
| ASE@MOE | PLI-CASF2016/CPSet |
| ASP@GOLD | PLI-CASF2016/CPSet |
| Affinity-dG@MOE | PLI-CASF2016/CPSet |
| Alpha-HB@MOE | PLI-CASF2016/CPSet |
| ChemPLP@GOLD | PLI-CASF2016 |
| ChemScore@GOLD | PLI-CASF2016/CPSet |
| ChemScore@SYBYL | PLI-CASF2016 |
| D-Score@SYBYL | PLI-CASF2016 |
| DrugScore2018 | PLI-CASF2016 |
| DrugScoreCSD | PLI-CASF2016 |
| G-Score@SYBYL | PLI-CASF2016 |
| GBVI/WSA-dG@MOE | PLI-CASF2016/CPSet |
| GlideScore-SP | PLI-CASF2016/CPSet |
| GlideScore-XP | PLI-CASF2016/CPSet |
| GoldScore@GOLD | PLI-CASF2016/CPSet |
| Jain@DS | PLI-CASF2016 |
| LUDI1@DS | PLI-CASF2016 |
| LUDI2@DS | PLI-CASF2016 |
| LUDI3@DS | PLI-CASF2016 |
| LigScore1@DS | PLI-CASF2016 |
| LigScore2@DS | PLI-CASF2016 |
| London-dG@MOE | PLI-CASF2016/CPSet |
| PLP1@DS | PLI-CASF2016 |
| PLP2@DS | PLI-CASF2016 |
| PMF04@DS | PLI-CASF2016 |
| PMF@DS | PLI-CASF2016 |
| PMF@SYBYL | PLI-CASF2016 |
| X-Score | PLI-CASF2016 |
| X-ScoreHM | PLI-CASF2016 |
| X-ScoreHP | PLI-CASF2016 |
| X-ScoreHP | PLI-CASF2016 |
| X-ScoreHS | PLI-CASF2016 |
| EGNN | PNI，NLI，Protein–Macrocycle |
| SchNet | PNI，NLI，Protein–Macrocycle |
| LEFTNet | PNI，NLI，Protein–Macrocycle |
| TorchMD-Net | PNI，NLI，Protein–Macrocycle |
| DOCK6 | NLI |
| PLANTS | NLI |
| rDock | NLI-RNA-Ligand,PLI-CPSet |
| Surflex | NLI-RNA-Ligand,PLI-CPSet |
| MM/GBSA | NLI |
| Gold@plp | PLI-CPSet |
| Rosetta | PLI-CPSet |
| CSM-carbohydrate | PLI-PCAPRED |
| PCAPRED | PLI-PCAPRED |
| CSM-lig | PLI-PCAPRED |
| NNscore (pdbbind2016) | PLI-PCAPRED |
| RFscore (v3 pdbbind2016) | PLI-PCAPRED |
| PLEC score (pdbbind2016) | PLI-PCAPRED |

PLI: Protein–Ligand Interaction. PPI: Protein–Protein Interaction. PNI: Protein–Nucleic Acid Interaction. NLI: Nucleic Acid–Ligand Interaction. Protein–Macrocycle: Protein–Non-Peptidic Macrocycle Test. RNA-Ligand: The test subset includes only 29 RNA–small molecule complexes. For methods that are only involved in certain subsets of a specific interaction type, this will be explicitly indicated. For example: PLI-DUD-E/DEKOIS2.0. Descriptions of specific test sets can be found in Table S2. Some method results are obtained directly from the corresponding publications[1–6].

Table S2. Details of Evaluation Test Sets

| Dataset | Source | Type | Power | Processed |
|---|---|---|---|---|
| CASF-2016 | Ref[1] | PLI | scoring | 284 |
| | | | ranking | 284 |
| | | | docking | 22,760 |
| | | | screening | 1,622,628 |
| DEKOIS 2.0 | Ref[7] | PLI | screening | 867,020 |
| DUD-E | Ref[8] | PLI | screening | 11,903,802 |
| PPI Benchmark | Ours | PPI | scoring | 79 |
| | | | docking | 7,979 |
| | | | screening | 613,900 |
| SAbDab | Ref[9] | PPI | scoring | 150 |
| Peptide-MHC I | Ref[10] | PPI | screening | 42,454 |
| PNI test | Ref[11] | PNI | scoring | 554 |
| NLI test | Ref[11] | NLI | scoring | 134 |
| Protein–Non-Peptidic Macrocycle Test | Ours | PLI | scoring | 261 |
| CPSet | Ref[12] | PLI | scoring | 270 |
| PCAPRED | Ref[13] | PLI | scoring | 43 |

Table S3. Ablation Study Results on Protein–Protein Complexes

| Model | Docking – PPI Benchmark | | | | | | Screening – PPI Benchmark | | | Average Rank |
|---|---|---|---|---|---|---|---|---|---|---|
| | Spearman (All) | Spearman (Top 20) | SR (Top 1) | SR (Top 5) | SR (Top 10) | Hit (Top 1) | $SR_{1\%}$ | $SR_{5\%}$ | $SR_{10\%}$ | |
| Ablation 1 | 0.5617 | 0.4134 | 63.29% | 89.87% | 96.20% | 88.61% | 6.33% | 16.46% | 24.05% | 26.15% |
| Ablation 2 | -0.2361 | 0.3458 | 26.58% | 53.16% | 67.09% | 26.58% | 2.53% | 7.59% | 13.92% | 43.23 |
| Ablation 3 | 0.4083 | 0.5020 | 100.00% | 100.00% | 100.00% | 100.00% | 1.27% | 12.66% | 18.99% | 32.39 |
| BioScore | 0.6382 | 0.5545 | 100.00% | 100.00% | 100.00% | 100.00% | 15.19% | 32.91% | 41.77% | 21.54 |

Ablation 1: Without interface-masking encoding strategy; inter-molecular edges are introduced into the complex geometry graph. Ablation 2: Traditional statistical potential–based scoring strategy; uses the summation of statistical potentials as the final score. Ablation 3: Without interaction edge count–aware confidence score; final score is based solely on the mean of statistical potentials.

Table S4. Ablation Study Results on Protein–Small Molecule Complexes

| Model | Docking – CASF2016 | | | | | | Screening – CASF2016 | | | | | |
|---|---|---|---|---|---|---|---|---|---|---|---|---|
| | SR (Top 1) | SR (Top 5) | SR (Top 10) | Hit (Top 1) | Hit (Top 2) | Hit (Top 5) | $SR_{1\%}$ | $SR_{5\%}$ | $SR_{10\%}$ | $EF_{1\%}$ | $EF_{5\%}$ | $EF_{10\%}$ |
| Ablation 1 | 22.11% | 34.04% | 52.98% | 78.25% | 87.37% | 94.74% | 8.77% | 22.81% | 31.58% | 1.81 | 1.71 | 1.61 |
| Ablation 2 | 42.46% | 58.95% | 79.30% | 72.98% | 80.00% | 91.93% | 0.00% | 3.51% | 8.77% | 1.23 | 1.48 | 1.43 |
| Ablation 3 | 62.46% | 80.00% | 94.04% | 94.39% | 96.84% | 99.30% | 19.30% | 40.35% | 49.12% | 14.54 | 6.07 | 4.03 |
| BioScore | 65.26% | 82.46% | 96.14% | 96.14% | 97.89% | 99.30% | 68.42% | 80.70% | 89.47% | 29.11 | 9.95 | 5.96 |

Ablation 1: Without interface-masking encoding strategy; inter-molecular edges are introduced into the complex geometry graph. Ablation 2: Traditional statistical potential–based scoring strategy; uses the summation of statistical potentials as the final score. Ablation 3: Without interaction edge count–aware confidence score; final score is based solely on the mean of statistical potentials.

Table S5. Inference Time Statistics for Different Scoring Methods

| Method | Inference Time per Complex (s) |
|---|---|
| GET | 0.0156 |
| GNN-DOVE | 0.3283 |
| DProQA | 0.7681 |
| VoroMQA | 3.8648 |
| ZRANK2 | 0.0629 |
| MINT | 0.0001 |
| AutoDock Vina | 2.1300 |
| RTMScore | 0.1266 |
| BioScore | 0.0086 |

Table S6. Description of All Pretraining and Fine-Tuning Datasets for BioScore

| Dataset | Type | Raw | Period | Processed (train / valid) |
|---|---|---|---|---|
| PDBbind (version 2020) | PLI | 19,443 | pretrain | 16,223 / 1,500 |
| | | | fintune | 12,351 / 1,500 |
| | PPI | 2,852 | pretrain | 1,776 / 163 |
| | | | finetune | 1,776 / 163 |
| | PNI | 1,052 | pretrain | 276 / 92 |
| | | | finetune | 276 / 92 |
| SAbDab | PPI | 736 | finetune | 112 / 10 |
| CSM-carbohydrate | PLI | 327 | finetune | 327 / 0 |

Table S7. Scoring Performance Results for Antigen–Antibody Complexes

| Model | Pearson | Spearman | RMSE | MAE |
|---|---|---|---|---|
| GET | 0.3659 | 0.3354 | 1.9637 | 1.4849 |
| MINT | 0.0462 | 0.0434 | 1.6497 | 1.2592 |
| BioScore | 0.4628 | 0.4318 | 1.5133 | 1.2000 |

Table S8. CASF-2016 Benchmark Results on Protein–Small Molecule Complexes

| Model | Scoring | | | Ranking | | Docking | Screening | | | | | |
|---|---|---|---|---|---|---|---|---|---|---|---|---|
| | Pearson | RMSE | MAE | Spearman | PI | Hit (Top 1) | $SR_{1\%}$ | $SR_{5\%}$ | $SR_{10\%}$ | $EF_{1\%}$ | $EF_{5\%}$ | $EF_{10\%}$ |
| BioScore | 0.801 | 1.30 | 1.00 | 0.704 | 0.737 | 96.1% | 68.4% | 80.7% | 89.5% | 29.11 | 9.95 | 5.96 |
| PIGNet | 0.761 | - | - | 0.682 | - | 87.0% | 55.4% | - | - | 19.60 | - | - |
| DeepDock | 0.460 | - | - | 0.425 | - | 87.0% | 43.9% | - | - | 16.41 | - | - |
| RTMScore | 0.455 | - | - | 0.529 | - | 97.3% | 66.7% | - | - | 28.00 | - | - |
| GenScore | 0.773 | - | - | 0.659 | - | 97.6% | 71.4% | - | - | 28.16 | - | - |
| PIGNet2 | 0.747 | - | - | 0.651 | - | 93.0% | 66.7% | - | - | 24.90 | - | - |
| ASE@MOE | 0.591 | - | - | 0.439 | 0.466 | 50.5% | 7.0% | 12.3% | 28.1% | 1.44 | 1.11 | 1.28 |
| ASP@GOLD | 0.617 | - | - | 0.553 | 0.582 | 81.1% | 22.8% | 49.1% | 68.4% | 6.98 | 3.95 | 3.10 |
| Affinity-dG@MOE | 0.552 | - | - | 0.604 | 0.619 | 63.5% | 19.3% | 43.9% | 50.9% | 5.07 | 2.77 | 2.26 |
| Alpha-HB@MOE | 0.569 | - | - | 0.535 | 0.558 | 71.6% | 8.8% | 21.1% | 40.4% | 1.70 | 1.65 | 2.02 |
| Autodock Vina | 0.604 | - | - | 0.528 | 0.557 | 90.2% | 29.8% | 40.4% | 50.9% | 7.70 | 4.01 | 2.87 |
| ChemPLP@GOLD | 0.614 | - | - | 0.633 | 0.657 | 86.0% | 35.1% | 61.4% | 64.9% | 11.91 | 5.29 | 3.59 |
| ChemScore@GOLD | 0.574 | - | - | 0.526 | 0.558 | 80.4% | 28.1% | 45.6% | 57.9% | 8.65 | 3.95 | 2.92 |
| ChemScore@SYBYL | 0.59 | - | - | 0.593 | 0.617 | 57.9% | 1.8% | 15.8% | 31.6% | 0.79 | 1.26 | 1.41 |
| D-Score@SYBYL | 0.531 | - | - | 0.577 | 0.598 | 26.0% | 5.3% | 17.5% | 26.3% | 1.24 | 1.65 | 1.34 |
| DrugScore2018 | 0.602 | - | - | 0.607 | 0.637 | 83.5% | 15.8% | 31.6% | 38.6% | 3.66 | 2.25 | 1.89 |
| DrugScoreCSD | 0.596 | - | - | 0.63 | 0.663 | 87.4% | 22.8% | 33.3% | 49.1% | 5.90 | 2.97 | 2.54 |
| G-Score@SYBYL | 0.572 | - | - | 0.591 | 0.609 | 44.2% | 3.5% | 12.3% | 26.3% | 0.89 | 1.06 | 1.11 |
| GBVI/WSA-dG@MOE | 0.496 | - | - | 0.489 | 0.504 | 87.0% | 26.3% | 45.6% | 59.6% | 7.62 | 3.61 | 2.77 |
| GlideScore-SP | 0.513 | - | - | 0.419 | 0.425 | 87.7% | 36.8% | 54.4% | 63.2% | 11.44 | 5.83 | 3.98 |
| GlideScore-XP | 0.467 | - | - | 0.257 | 0.255 | 83.9% | 26.3% | 45.6% | 52.6% | 8.83 | 4.75 | 3.51 |
| GoldScore@GOLD | 0.416 | - | - | 0.284 | 0.283 | 75.1% | 15.8% | 35.1% | 42.1% | 4.27 | 2.86 | 1.98 |
| Jain@DS | 0.457 | - | - | 0.521 | 0.545 | 55.8% | 7.0% | 15.8% | 33.3% | 1.35 | 1.31 | 1.56 |
| LUDI1@DS | 0.494 | - | - | 0.612 | 0.64 | 63.2% | 14.0% | 29.8% | 42.1% | 3.10 | 2.14 | 1.81 |
| LUDI2@DS | 0.526 | - | - | 0.629 | 0.657 | 63.5% | 10.5% | 28.1% | 40.4% | 2.34 | 2.00 | 1.63 |
| LUDI3@DS | 0.502 | - | - | 0.532 | 0.564 | 53.0% | 7.0% | 14.0% | 33.3% | 1.85 | 1.15 | 1.40 |
| LigScore1@DS | 0.425 | - | - | 0.599 | 0.606 | 76.8% | 22.8% | 36.8% | 49.1% | 6.32 | 3.68 | 2.74 |
| LigScore2@DS | 0.54 | - | - | 0.608 | 0.62 | 85.6% | 26.3% | 42.1% | 50.9% | 6.82 | 3.53 | 2.84 |
| London-dG@MOE | 0.405 | - | - | 0.593 | 0.609 | 63.2% | 7.0% | 26.3% | 42.1% | 2.05 | 2.30 | 2.09 |
| PLP1@DS | - | - | - | 0.582 | 0.605 | 82.8% | 15.8% | 31.6% | 45.6% | 3.98 | 2.88 | 2.39 |
| PLP2@DS | 0.563 | - | - | 0.589 | 0.617 | 79.3% | 8.8% | 29.8% | 52.6% | 1.81 | 2.43 | 2.49 |
| PMF04@DS | 0.212 | - | - | 0.481 | 0.497 | 46.3% | 14.0% | 19.3% | 33.3% | 3.17 | 1.68 | 1.75 |
| PMF@DS | 0.422 | - | - | 0.537 | 0.559 | 42.8% | 14.0% | 26.3% | 40.4% | 3.76 | 1.76 | 1.59 |
| PMF@SYBYL | 0.262 | - | - | 0.449 | 0.478 | 47.7% | 7.0% | 19.3% | 28.1% | 1.46 | 1.77 | 1.68 |
| X-Score | 0.631 | - | - | 0.604 | 0.638 | 63.5% | 7.0% | 15.8% | 28.1% | 2.68 | 1.31 | 1.23 |
| X-ScoreHM | 0.609 | - | - | 0.603 | 0.641 | 65.3% | 8.8% | 19.3% | 31.6% | 3.21 | 1.39 | 1.31 |
| X-ScoreHP | 0.621 | - | - | - | - | - | - | - | - | - | - | - |
| X-ScoreHP | - | - | - | 0.573 | 0.607 | 56.1% | 3.5% | 17.5% | 29.8% | 1.79 | 1.54 | 1.13 |
| X-ScoreHS | 0.629 | - | - | 0.547 | 0.577 | 59.6% | 5.3% | 12.3% | 28.1% | 2.17 | 1.26 | 1.26 |
| ΔVinaRF20 | 0.625 | - | - | 0.588 | 0.612 | 30.2% | 5.3% | 14.0% | 24.6% | 1.76 | 1.12 | 1.15 |

Baseline model results are sourced from the corresponding publications[1,4,5,14–16].

Table S9. Controlled Comparison on the PPI Benchmark

| Model | Scoring | | | Docking | | | | | | Screening | | | |
|---|---|---|---|---|---|---|---|---|---|---|---|---|---|
| | Pearson | RMSE | MAE | Spearman (All) | Spearman (Top 20) | SR (Top 1) | SR (Top 5) | SR (Top 10) | Hit (Top 1) | $SR_{1\%}$ | $SR_{5\%}$ | $SR_{10\%}$ | Average Rank |
| BioScore-PPI | 0.22 | 2.43 | 1.88 | 0.67 | 0.56 | 97.47% | 98.73% | 100% | 98.73% | 8.86% | 21.52% | 31.65% | 24.56 |
| BioScore-PPI+PLI | 0.46 | 2.24 | 1.65 | 0.64 | 0.55 | 100.00% | 100.00% | 100.00% | 100.00% | 15.19% | 32.91% | 41.77% | 21.54 |

BioScore-PPI indicates that only PPI data were used during both pretraining and fine-tuning. BioScore-PPI+PLI indicates that mixed PLI and PPI data were used during pretraining, and PPI data were used during fine-tuning.

Table S10. Controlled Comparison on the SAbDab Test Set

| Model | Scoring | | |
|---|---|---|---|
| | Pearson | RMSE | MAE |
| BioScore-PPI | 0.24 | 1.66 | 1.35 |
| BioScore-PPI+PLI | 0.46 | 1.51 | 1.20 |

BioScore-PPI refers to the setting where only PPI data were used during pretraining, while BioScore-PPI+PLI denotes the use of mixed PPI and PLI data during pretraining. To avoid data leakage, both models were fine-tuned on the SAbDab training set that had been pre-partitioned specifically for leakage prevention.

Table S11. Controlled Comparison on the CASF-2016 Benchmark

| Model | Scoring | | | Ranking | | Docking | Screening | | | | | |
|---|---|---|---|---|---|---|---|---|---|---|---|---|
| | Pearson | RMSE | MAE | Spearman | PI | Hit (Top 1) | $SR_{1\%}$ | $SR_{5\%}$ | $SR_{10\%}$ | $EF_{1\%}$ | $EF_{5\%}$ | $EF_{10\%}$ |
| BioScore-PLI | 0.80 | 1.29 | 0.96 | 0.56 | 0.68 | 95.44% | 61.40% | 85.96% | 92.98% | 28.98 | 10.32 | 6.07 |
| BioScore-PPI+PLI | 0.80 | 1.30 | 1.00 | 0.70 | 0.74 | 96.14% | 68.42% | 80.70% | 89.47% | 29.11 | 9.95 | 5.96 |

BioScore-PLI indicates that only PLI data were used during both pretraining and fine-tuning. BioScore-PPI+PLI indicates that mixed PLI and PPI data were used during pretraining, and PLI data were used during fine-tuning.

Table S12. CPSet Benchmark Results

| Model | Scoring | | |
|---|---|---|---|
| | Pearson | RMSE | MAE |
| BioScore-PLI | 0.52 | 1.50 | 1.23 |
| BioScore-PPI | 0.24 | 1.77 | 1.44 |
| BioScore-PPI+PLI | 0.60 | 1.41 | 1.11 |

BioScore-PPI indicates that only PPI data were used during both pretraining and fine-tuning. BioScore-PLI indicates that only PLI data were used during both pretraining and fine-tuning. BioScore-PPI+PLI indicates that mixed PLI and PPI data were used during pretraining, and PPI data were used during fine-tuning.

Table S13. Test Results of Different Fine-Tuning Strategies Across Multiple Complex Systems

| Model | PPI-Scoring | | | PLI-Scoring | | | PLI-Ranking | | CPSet-Scoring | | |
|---|---|---|---|---|---|---|---|---|---|---|---|
| | Pearson | RMSE | MAE | Pearson | RMSE | MAE | Spearman | PI | Pearson | RMSE | MAE |
| BioScore-Finetune-mix | 0.32 | 2.10 | 1.60 | 0.79 | 1.35 | 1.01 | 0.64 | 0.65 | 0.52 | 1.58 | 1.24 |
| BioScore-Finetune-specific | 0.46 | 2.24 | 1.65 | 0.80 | 1.30 | 1.00 | 0.70 | 0.74 | 0.60 | 1.41 | 1.11 |

BioScore-Finetune-mix indicates that mixed PLI and PPI data were used during fine-tuning. BioScore-Finetune-specific indicates that only domain-specific data were used during fine-tuning. Both strategies use mixed PLI and PPI data during pretraining.

Table S14. PNI Scoring Test Set Results

| Model | Pearson | RMSE | MAE |
|---|---|---|---|
| GET | -0.0641 | 7.0047 | 3.9051 |
| EGNN | 0.1484 | 2.1838 | 1.6863 |
| SchNet | 0.2201 | 2.2666 | 1.7683 |
| LEFTNet | 0.0077 | 3.6586 | 2.6018 |
| TorchMD-Net | 0.1859 | 2.4497 | 1.9066 |
| BioScore | 0.3761 | 1.5753 | 1.2434 |

Table S15. NLI Scoring Test Set Results

| Model | Pearson | RMSE | MAE |
|---|---|---|---|
| GET | 0.3589 | 3.2319 | 2.7872 |
| EGNN | 0.3901 | 2.2046 | 1.7716 |
| SchNet | 0.3673 | 2.4325 | 2.0900 |
| LEFTNet | 0.3400 | 2.9245 | 2.4267 |
| TorchMD-Net | 0.2134 | 3.0295 | 2.5673 |
| BioScore | 0.4504 | 1.9070 | 1.5965 |

Table S16. Test Results on Protein–Non-Peptidic Macrocycle Complexes

| Model | Pearson | Spearman | RMSE | MAE |
|---|---|---|---|---|
| GET | 0.5245 | 0.5360 | 1.5902 | 1.3091 |
| EGNN | 0.5069 | 0.5264 | 1.4761 | 1.2154 |
| SchNet | 0.5266 | 0.5198 | 1.4790 | 1.2343 |
| LEFTNet | 0.4749 | 0.4818 | 1.4824 | 1.2172 |
| TorchMD-Net | 0.5451 | 0.5605 | 1.5064 | 1.2239 |
| AutoDock Vina | 0.4560 | 0.4667 | 2.1260 | 1.6621 |
| BioScore | 0.6195 | 0.6376 | 1.3638 | 1.0875 |

Table S17. Test Results on Protein–Carbohydrate Complexes

| Model | Pearson | Spearman | RMSE | MAE |
| --- | --- | --- | --- | --- |
| CSM-carbohydrate | 0.67 | 0.64 | 1.72 | 1.29 |
| PCAPRED | 0.46 | 0.58 | 2.73 | 1.79 |
| CSM-lig | 0.20 | 0.14 | 5.07 | 4.04 |
| NNscore (pdbbind2016) | 0.40 | 0.36 | 2.49 | 2.01 |
| RFscore (v3 pdbbind2016) | 0.36 | 0.33 | 2.19 | 1.78 |
| PLEC score (pdbbind2016) | 0.49 | 0.56 | 2.49 | 1.77 |
| Autodock Vina | 0.28 | 0.29 | 2.54 | 1.86 |
| BioScore | 0.76 | 0.72 | 1.09 | 0.79 |

Baseline model results are sourced from the work by Nguyen et al.[3]

Table S18. BioScore Hyperparameter Settings

| Hyperparameters | Setting |
| --- | --- |
| Learning rate (Pretrain) | $10^{-3}$ |
| Learning rate (Finetune) | $10^{-4}$ |
| Weight decay (Pretrain) | $10^{-8}$ |
| Weight decay (Finetune) | $10^{-5}$ |
| Maximum number of epochs | 200 |
| Patience of early stopping | 10 |
| Hidden dimension of GET layer | 128 |
| Upperbound of the number of nodes in a batch | 2500 |
| Number of GET layers | 3 |
| Number of attention heads | 4 |

## 2. Supplementary Figures

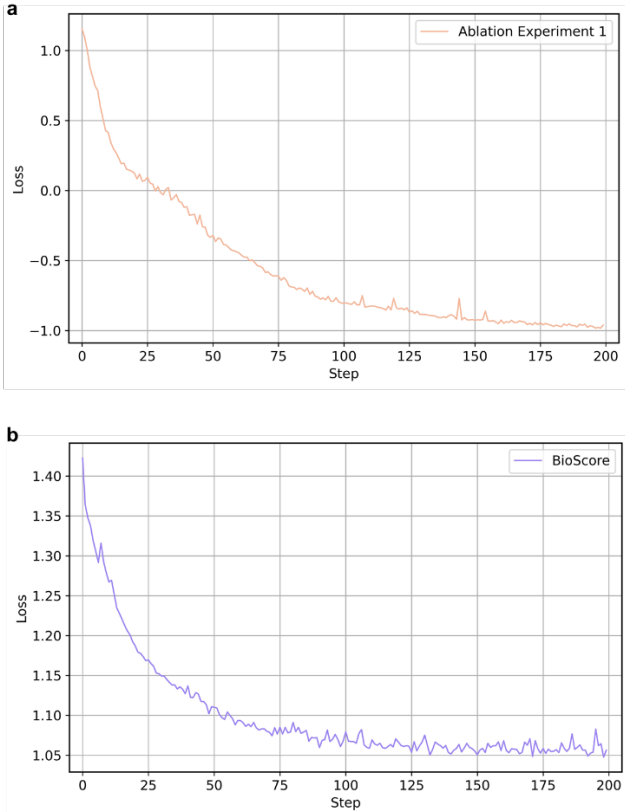

Figure S1. Validation Curves for Ablation Experiment 1 (without interface-masking encoding) and BioScore (with interface-masking encoding) during Mixed Pretraining on PLI and PPI Data.

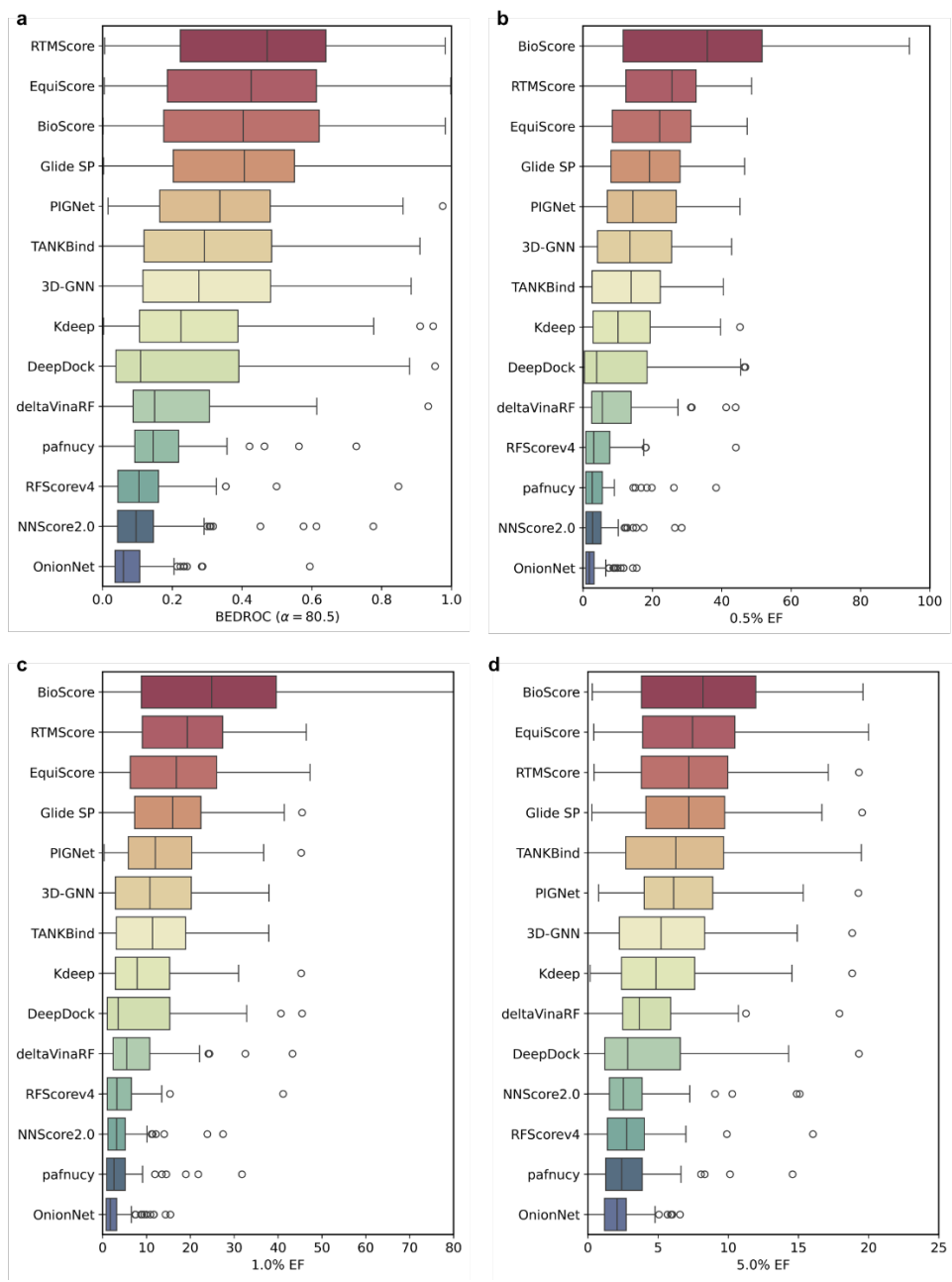

Figure S2. Evaluation of 14 Scoring Methods on DUD-E. (A–D) Ranking of all methods by mean values for BEDROC (α=80.5), 0.5% EF, 1.0% EF, and 5.0% EF, respectively.

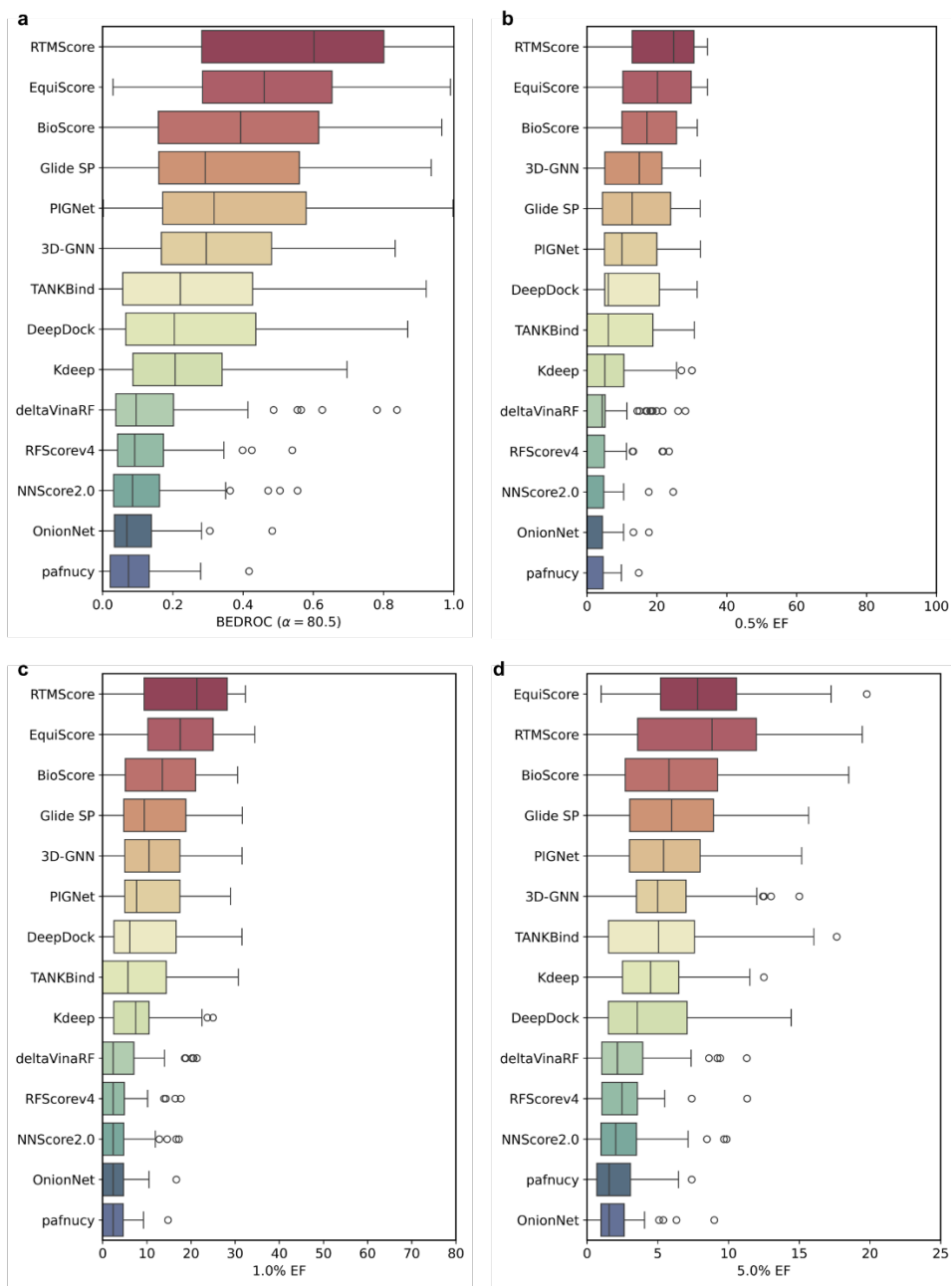

Figure S3. Evaluation of 14 Scoring Methods on DEKOIS2.0. (A–D) Ranking of all methods by mean values for BEDROC (α=80.5), 0.5% EF, 1.0% EF, and 5.0% EF, respectively.

## 3. PPI Benchmark

To construct a comprehensive benchmark for scoring functions targeting protein–protein complexes, we performed an initial filtering of complexes from PDBbind (version 2020) based on the following criteria: (1) the protein–protein interaction structure must consist of two heteromeric chains, with a sequence similarity between the chains of no more than 30%; (2) all structures must be non-NMR resolved, with a resolution better than 3.5 Å; (3) no non-protein macromolecular polymers (such as nucleic acids) may be present; (4) no missing amino acids are allowed in the structures; (5) both chains must have fewer than 500 residues; and (6) a clear binding affinity label must be available, with the type being "Kd." As a result, we selected 177 protein–protein complexes from 2,852 entries in the PDBbind protein–protein dataset that met these criteria.

Next, we categorized the protein chains within the complexes based on sequence length into three groups: A (<50 amino acids), B (50–250 amino acids), and C (250–500 amino acids). Based on the lengths of the interacting heteromeric chains (i.e., receptor and ligand chains), we further divided the complexes into four groups: A-B, A-C, B-B, and B-C. We then randomly selected 20 PPIs from each group, substituting entries with others from the same group if necessary to meet the downstream benchmark construction requirements (notably, for the A-C group, only 19 complexes met the criteria for constructing the protein–protein docking benchmark). Following this process, a final set of 79 high-quality protein–protein complexes with diverse sequence length distributions was curated and designated as the foundational complex dataset for our PPI benchmark.

To comprehensively evaluate the different capabilities of scoring functions, we followed the design principles of CASF-2016[1]. We first constructed a scoring power benchmark by using the structures and corresponding experimental binding affinity annotations from the foundational complex dataset. This benchmark evaluates the ability of a scoring function to predict the binding affinity of protein–protein complexes. It is important to note that, due to the lack of sufficient data in PDBbind where a single protein receptor is associated with multiple protein ligands, it is not feasible to construct a ranking power benchmark for the PPI dataset similar to that of CASF-2016.

We further constructed a docking power benchmark based on the foundational complex dataset to evaluate the ability of scoring functions to distinguish native-like from non-native binding conformations. We used ZDOCK 3.0.2[17] to generate decoy conformations for the 79 heterodimeric targets. In the decoy generation and sampling process, the longer chain was designated as the receptor and the shorter chain as the ligand. Using a 6-degree Euler angle increment, each ZDOCK run produced 6,000 models per complex, and each complex underwent five independent ZDOCK runs. For each run, the monomers were randomly rotated and translated in space to ensure all generated decoy models were distinct. For each target, we computed DockQ scores for all 30,000 models and binned the docking poses into 20 intervals based on their DockQ scores (ranging from 0 to 1) with a bin width of 0.05. For each target, we selected five models from each DockQ bin, following the principle of maximizing DockQ diversity. Due to the scarcity of decoy models with DockQ > 0.9, we combined the bins for 0.9 < DockQ < 0.95 and DockQ > 0.95 into a single bin, from which 10 models were selected. This procedure yielded 100 decoy models per target. It is important to note that some complexes did not have sufficient decoy models in every DockQ bin to meet the sampling requirements. These PPIs were replaced with entries from the same sequence length group, with corresponding substitutions made in the protein–protein foundational dataset to maintain the completeness and consistency of the PPI benchmark. As a result, the A-C group in the final foundational dataset contains only 19 complexes (the number of complexes that met the criteria).

Finally, we further performed cross-docking of the 79 receptor proteins and their partner proteins in the foundational complex dataset using ZDOCK. Given the potentially large number of receptor–ligand pairs, we reduced the number of docking poses generated by ZDOCK for each ligand to 600. Using the K-means clustering method, we grouped the 600 docking poses based on ligand conformational similarity, measured by Root Mean Square Deviation (RMSD) values, into 100 clusters. From each cluster, we randomly selected one representative pose to ensure that each receptor–ligand pair's decoy dataset contained 100 diverse binding conformations. In the post-processing step, we standardized chain identifiers by renaming receptor chains as "A" and ligand chains as "B" to avoid duplicate native chain IDs during cross-docking. Importantly, to mitigate the risk of false negatives in the cross-docking evaluation, we performed 90% sequence identity clustering of receptor chains using CD-HIT and treated ligands within the same cluster as potential false-negative ligands, which were excluded from the screening evaluation. This process yielded a screening power benchmark comprising 613,900 docking conformations for assessing a scoring function's ability to distinguish active from inactive ligands.

Final Selected 79 PDB IDs and Groupings

Total

['6f0f', '1j2j', '1mzw', '5yip', '4js0', '3h8k', '2vay', '5wuj', '6j4s', '6e3j', '2wh6', '5gtb', '6jjw', '5h3j', '5inb', '2gww', '5h9b', '3ukz', '5ky4', '2qxv', '6bw9', '4uyq', '2hth', '3ul4', '1gua', '4dbg', '3u43', '3fpu', '1kac', '6ne4', '2v9t', '5ml9', '2uyz', '4eig', '1lw6', '3vv2', '3zu7', '5jw7', '5g15', '2f4m', '4z9k', '4d0n', '6hul', '1tdq', '2qc1', '2v8s', '2omz', '2voh', '3gj6', '3kj0', '3tz1', '4d0g', '4rey', '4y61', '1pjm', '1pjn', '1syq', '1rkc', '1t01', '3ukx', '3ul0', '3ul1', '5ky5', '6iu7', '6iua', '6k06', '1emv', '1p69', '2vlp', '2vln', '3kuc', '2sic', '1jiw', '2b7c', '2omw', '2v3b', '2z58', '3vyr', '4m0w']

Group: A-B

['1j2j', '1mzw', '2voh', '2vay', '3h8k', '3gj6', '2wh6', '3kj0', '3tz1', '4js0', '4d0g', '4rey', '5h3j', '5wuj', '5gtb', '6e3j', '5yip', '6j4s', '6f0f', '6jjw']

Group: A-C

['1pjm', '1pjn', '1syq', '1rkc', '1t01', '2gww', '2qxv', '3ukx', '3ul0', '3ukz', '3ul1', '5inb', '5ky5', '5ky4', '5h9b', '6bw9', '6iu7', '6iua', '6k06']

Group: B-B

['1gua', '1kac', '1emv', '1p69', '2hth', '2qc1', '2v8s', '2v9t', '2uyz', '2vlp', '2vln', '3fpu', '3kuc', '4dbg', '3ul4', '3u43', '4eig', '4uyq', '5ml9', '6ne4']

Group: B-C

['2sic', '1jiw', '1lw6', '1tdq', '2b7c', '2f4m', '2omw', '2omz', '2v3b', '2z58', '3vyr', '3zu7', '3vv2', '4d0n', '4m0w', '4y61', '4z9k', '5jw7', '5g15', '6hul']

Ligands within the same cluster are considered potential false-negative ligands and are excluded from the screening evaluation.

Cluster 3 ['2omz', '2omw']; Cluster 7 ['3ul1', '3ukz', '6iu7', '6iua', '1pjn', '3ul0', '1pjm', '3ukx', '6k06']; Cluster 18 ['5ky4', '5ky5']; Cluster 21 ['3vv2', '2z58']; Cluster 35 ['1lw6', '2sic']; Cluster 43 ['2gww', '1syq', '1t01', '1rkc']; Cluster 67 ['1kac', '1p69']; Cluster 78 ['1gua', '3kuc']; Cluster 110 ['1emv', '2vln', '2vlp'].

# 4. Basic Evaluation Methods

Following CASF-2016, there are four different tasks assessed in our evaluation, including the scoring, docking, ranking, and screening powers. Here, we introduce the primary assessment indicators for each of the four powers in our evaluations.

"Scoring power" refers to the ability of a scoring function to produce binding scores in a linear correlation with experimental binding data. The Pearson's correlation coefficient between the computed binding scores and the experimental binding constants was calculated as a quantitative measure of the scoring power. Root Mean Square Error (RMSE) and Mean Absolute Error (MAE) were also used to evaluate the predictive accuracy of the scoring function.

"Ranking power" refers to the ability of a scoring function to correctly rank the known ligands of a certain target by their binding affinities when the precise binding poses of those ligands are given. The Spearman's rank correlation coefficient and Predictive Index (PI) were used as the main quantitative indicators of ranking power.

"Docking power" refers to the ability of a scoring function to identify the native ligand binding pose among in-silico-generated decoys. Spearman's rank correlation coefficient is one of the most commonly used metrics for evaluating docking power. "Spearman's top-k" represents the ability of the scoring function to rank decoys within the top k based on their dockQ scores, it evaluates the scoring function's capability to identify higher-quality conformations within a set of high-quality but slightly different decoys. Success rate(SR) and hit rate(Hit) are two additional key indicators for evaluating docking power. The success rate refers to the percentage of successful predictions that the native pose ranked within the top k of all scored poses, while the hit rate is a metric where the complex can be marked as a successful prediction if one of the RMSD values between the best-scored poses and the native pose is below the predefined threshold (usually 2.0Å), for the PPI system, the DockQ score (with a threshold of 0.8) is used as the metric.

"Screening power" refers to the ability of a scoring function to identify the true binders to a given target among a pool of random ligands, and was evaluated essentially in a cross-docking trial. The indicator of the screening power includes the success rate (SR) of identifying the highest-affinity binder among the k% top-ranked ligands, and the enrichment factor (EF) which is defined as the percentage of true binders observed among all of the true binders for a given percentile of the top-ranked candidates. We also considered the ranking of the true ligand with the highest affinity among all ligands, using the Average Rank to represent the average performance across all targets for intuitive presentation. A higher rank indicates better ability of the scoring function to identify the true ligand. In addition, Area Under the Receiver Operating Characteristic Curve (AUROC), Area Under the Precision-Recall Curve (AUPR), and Boltzmann-Enhanced Discrimination of Receiver Operating Characteristic (BEDROC) were also considered to assess the screening power of scoring functions, with AUROC providing an overall measure of discriminative power, AUPR offering a more focused evaluation on how well the function performs in predicting relevant hits, especially when the positive class is scarce, and BEDROC providing a further evaluation that emphasizes early enrichment of true positives, which is crucial for prioritizing the most relevant candidates in screening.

# 5. Description of Baseline Methods

(1) GET

We obtained results of GET using the open-source codes at https://github.com/thunlp-mt/get and retrained with default settings. Following GET, basic baselines such as EGNN, TorchMD-Net, and LEFTNet were obtained from their respective open-source codes, and SchNet was implemented using PyTorch Geometric. The retraining procedures and data types for the aforementioned models align with the specifications outlined in the original GET methodology. For the PLI and PPI tests, a hybrid dataset comprising both PLI and PPI data was utilized for training. In tasks related to nucleic acids, a combined training approach incorporating PLI, PPI, and PNI datasets was employed.

(2) GNN-DOVE

We obtained results of GNN-DOVE using the open-source codes and weights at https://github.com/kiharalab/GNN_DOVE with default settings.

(3) DProQA

We obtained results of DProQA using the open-source codes and weights at https://github.com/jianlin-cheng/DProQA with default settings.

(4) MINT

We obtained the embeddings of MINT using the open-source codes and pretrained weights at https://github.com/VarunUllanat/mint with default settings, and then finetuned the downstream MLP to predict the binding affinity of protein complexes.

(5) VoroMQA

Software version: Voronota 1.22.3149-1. We applied the default settings of voronota-voromqa to score the quality of complex structures.

(6) ZRANK2

Software version: ZRANK 2.0, Pymol 2.4.1. Complex preparation: Hydrogen atoms were added with Pymol 2.4.1 and new complex PBD files with Hydrogen atoms were generated as input of ZRANK2. We scored the complexes by ZRANK2 without the parameter "-R" for initial-stage docked models.

(7) Autodock Vina

Software version: AutoDock Vina 1.2.3, MGLTOOLS 1.5.7. Ligand preparation: The ligands are processed into PDBQT files with hydrogen atoms added using the MGLTOOLS prepare_ligand4 scripts. Protein preparation: The proteins were processed into PDBQT files with hydrogen atoms added using the MGLTOOLS prepare_receptor4 scripts. The score_only mode was used to score the binding affinity of complexes without changing the pose.

The evaluation results of other scoring tools mentioned in this paper are based on data from the corresponding publications.